\documentclass[useAMS,usenatbib]{mn2e}

\usepackage{amssymb}
\usepackage{times}
\usepackage{graphicx}
\usepackage{epstopdf}
\usepackage{subfigure}
\usepackage[T1]{fontenc}
\usepackage{aecompl} 
\usepackage{multirow}
\usepackage{pdflscape}
\usepackage{rotating}

\usepackage[fleqn]{amsmath}
\usepackage{mathrsfs}
\usepackage{xcolor}

\usepackage{anyfontsize}

\usepackage{hyperref}

\newcommand{\beq}{\begin{equation}}
\newcommand{\eeq}{\end{equation}}

\def\gs{\mathrel{\lower0.6ex\hbox{$\buildrel {\textstyle >}\over{\scriptstyle \sim}$}}}
\def\ls{\mathrel{\lower0.6ex\hbox{$\buildrel {\textstyle <}\over{\scriptstyle \sim}$}}}
\newcommand{\simgt}{\lower.5ex\hbox{$\; \buildrel > \over \sim \;$}}
\newcommand{\simlt}{\lower.5ex\hbox{$\; \buildrel < \over \sim \;$}}

\newcommand{\aap}{A\&A}
\newcommand{\araa}{ARA\&A}
\newcommand{\apj}{ApJ}
\newcommand{\apjl}{ApJ}
\newcommand{\apjs}{ApJS}
\newcommand{\aj}{AJ}

\newcommand{\jcap}{J. Cosmol. Astropart. Phys.}
\newcommand{\pasj}{PASJ}
\newcommand{\prd}{Phys. Rev. D}
\newcommand{\mnras}{MNRAS}

\newcommand{\ssr}{Space Science Reviews}

\defcitealias{se+et14_comalit_I}{CoMaLit-I}
\defcitealias{ser+al14_comalit_II}{CoMaLit-II}
\defcitealias{ser14_comalit_III}{CoMaLit-III}

\begin{document}

\title[CoMaLit-IV]{CoMaLit -- IV. Evolution and self-similarity of scaling relations with the galaxy cluster mass
}
\author[
M. Sereno, S. Ettori
]{
Mauro Sereno$^{1,2}$\thanks{E-mail: \href{mailto:mauro.sereno@unibo.it}{mauro.sereno@unibo.it} (MS)}, Stefano Ettori$^{2,3}$
\\
$^1$Dipartimento di Fisica e Astronomia, Alma Mater Studiorum -- Universit\`a di Bologna, viale Berti Pichat 6/2, 40127 Bologna, Italia\\
$^2$INAF, Osservatorio Astronomico di Bologna, via Ranzani 1, 40127 Bologna, Italia\\
$^3$INFN, Sezione di Bologna, viale Berti Pichat 6/2, 40127 Bologna, Italia\\
}


\maketitle

\begin{abstract}
The scaling of observable properties of galaxy clusters with mass evolves with time. Assessing the role of the evolution is crucial to study the formation and evolution of massive halos and to avoid biases in the calibration. We present a general method to infer the mass and the redshift dependence, and the time-evolving intrinsic scatter of the mass--observable relations. The procedure self-calibrates the redshift dependent completeness function of the sample. The intrinsic scatter in the mass estimates used to calibrate the relation is considered too. We apply the method to the scaling of mass $M_\Delta$ versus line of sight galaxy velocity dispersion $\sigma_\mathrm{v}$, optical richness, X-ray luminosity, $L_\mathrm{X}$, and Sunyaev-Zel'dovich signal. Masses were calibrated with weak lensing measurements. The measured relations are in good agreement with time and mass dependencies predicted in the self-similar scenario of structure formation. The lone exception is the $L_\mathrm{X}$-$M_\Delta$ relation whose time evolution is negative in agreement with formation scenarios with additional radiative cooling and uniform preheating at high redshift. The intrinsic scatter in the $\sigma_\mathrm{v}$-$M_\Delta$ relation is notably small, of order of 14 per cent. Robust predictions on the observed properties of the galaxy clusters in the CLASH sample are provided as cases of study.
Catalogs and scripts are publicly available at \url{http://pico.bo.astro.it/\textasciitilde sereno/CoMaLit/}.
\end{abstract}


\section{Introduction}

Scaling relations among cluster global properties embody important clues on the formation and evolution of cosmic structures. They result from the main gravitational processes driving the cluster evolution \citep{kai86,gio+al13,bat+al12,ett13}. Accurate mass--observable relations are also needed to use the abundance of galaxy clusters to constrain cosmological parameters \citep{vik+al09,man+al10,man+al14,roz+al10,planck_2013_XX}.

This paper is the fourth in a series titled `CoMaLit' (COmparing MAsses in LITerature), which aims to assess our present capability to measure cluster masses, and to develop methods to measure scaling relations through Bayesian techniques. In the first paper \citep[ CoMaLit-I]{se+et14_comalit_I}, we evaluated systematic differences in lensing and X-ray masses obtained from independent analyses and we quantified the overall level of bias and intrinsic scatter of these mass proxies. The second paper presented the formalism to calibrate an observable cluster property against the cluster mass and applied the methodology to the Sunyaev-Zel'dovich (SZ) {\it Planck} selected clusters \citep[ CoMaLit-II]{ser+al14_comalit_II}. The Literature Catalogs of weak Lensing Clusters (LC$^2$), which are standardised and homogenised compilations of clusters and groups with weak lensing (WL) mass estimates, were presented in the third paper \citep[ CoMaLit-III]{ser14_comalit_III}. Here, we extend the Bayesian approach to account for redshift evolution of the scaling relations, of the intrinsic scatters in the mass and in the observable, and of the selection function.

If gravity is the dominant process, the resulting self-similar model predicts scaling relations in form of scale-free power laws \citep{kai86,gio+al13}. Numerical simulations \citep{sta+al10,fab+al11} confirmed these scalings and showed that intrinsic scatters around the median relations approximately follow a log-normal distribution. This basic theoretical scheme is very successful in describing observed scaling relations in X-ray and SZ \citep{ett13,ett15}. Deviations from the self-similar scheme may indicate that non-gravitational processes, such as feedback and non-thermal processes, contribute significantly to the global energy budget in clusters \citep{mau+al12}.
 
The precise measurement of the redshift evolution of the scaling relations is then crucial to understand how either gravitational or non-gravitational phenomena drive the formation and evolution of clusters. Furthermore, if the time evolution is neglected or wrongly shaped, the estimated scaling with mass may be biased in cluster samples spanning a significant redshift range \citep{and14}.

Any real time-dependence in the scaling of observables with mass and redshift has to be separated by other effects connected either to the evolution of the cluster mass function or to the redshift dependence of the selection function and of the completeness of the sample. Further complications are due to the fact that usual samples of clusters are not assembled according to well defined criteria but they may be just heterogeneous collections of systems with high quality data. In this case, the determination of the selection function is very problematic.

Here we develop a method that measures at the same time the evolution of the scaling relation and the completeness/selection function of the sample. This is a self-calibrating method which is intended to be optimised in large optical survey such as Euclid \citep{eucl_lau_11}. If we calibrate a cluster observable against the cluster mass, this relation can be used to construct a mass proxy based on the observable. The optimal mass proxy is expected to be easy to measure, unbiased, and minimally scattered. A crucial aspect is that in the first step we cannot calibrate the observable against the true mass (which cannot be measured), but we have to rely on another mass proxy, such as the WL mass or the hydrostatic mass, which are scattered too (\citealt{ras+al12};\citetalias{se+et14_comalit_I}). This scatter has to be considered to avoid biases (\citealt{man+al14};\citetalias{se+et14_comalit_I,ser+al14_comalit_II})

We apply the method to calibrate mass proxies based on the line of sight velocity dispersion of cluster galaxies, which is supposedly the best mass proxy \citep{sta+al10,sar+al13}, and three other observables, which are more scattered but optimised for large surveys, i.e., the optical richness, the X-ray luminosity, and the Sunyaev-Zel'dovich integrated Compton parameter.

The paper is organised as follows. Section~\ref{sec_evol} is devoted to general considerations on the redshift evolution of the scaling relations, of the intrinsic scatters, and of the selection/completeness function. Section~\ref{sec_regr} reviews the methodology employed to perform the regression and to recover at the same time the scaling relations and the completeness and selection functions. The cluster catalogs used in the analysis are introduced in Sec.~\ref{sec_cata}. Results are presented in Sec.~\ref{sec_resu} whereas Sec.~\ref{sec_comparisons} is devoted to the comparison with theoretical predictions and previous works. Final considerations are contained in the Sec.~\ref{sec_conc}. In App.~\ref{app_mass}, we discuss how the masses of clusters in selected samples are usually distributed. Appendix~\ref{app_SC} describe the format of the compiled catalogs of line-of-sight velocity dispersions.

Throughout the CoMaLit series of papers, we have been adopting the following conventions and notations. The frame-work cosmological model is the concordance flat $\Lambda$CDM universe with density parameter $\Omega_\mathrm{M}=0.3$, and Hubble constant $H_0=70~\mathrm{km~s}^{-1}\mathrm{Mpc}^{-1}$. $H(z)$ is the redshift dependent Hubble parameter and $E_z\equiv H(z)/H_0$. When $H_0$ is not specified, $h$ is the Hubble constant in units of $100~\mathrm{km~s}^{-1}\mathrm{Mpc}^{-1}$. 

$O_{\Delta}$ denotes a global property of the cluster measured within the radius which encloses a mean over-density of $\Delta$ times the critical density at the cluster redshift, $\rho_\mathrm{cr}=3H(z)^2/(8\pi G)$.  `$\log$' is the logarithm to base 10 and `$\ln$' is the natural logarithm.

\section{Redshift evolution}
\label{sec_evol}

Scaling relations evolve with redshift. Numerical simulations \citep{sta+al10} and theoretical predictions \citep{gio+al13} agree that the relation between the mass $M_\Delta$ and any observable quantity $O$ can be summarised by the form
\beq
\label{eq_evo_1}
O \propto M_\Delta^\beta E_z^\gamma.
\eeq
Within this framework, the redshift evolution in the median scaling relation is accounted for by the factor $E_z^\gamma$ whereas the slope $\beta$ is redshift independent. In fact, in the self-similar scenario, the evolution does not depend on the mass scale and only the normalisation depends on cosmic time. 

We tested this scheme in a number of cases. We considered observables connected either to the galaxy distribution, i.e., velocity dispersion of galaxies along the line of sight, $\sigma_\mathrm{v}$, or optical richness, $\lambda$, or to the intra-cluster medium, i.e., the bolometric X-ray luminosity, $L_\mathrm{X}$, or the spherically integrated SZ Compton signal, $Y_\mathrm{SZ}$. In the self-similar scenario, we expect that for clusters in equilibrium the scalings go like \citep{gio+al13,ett15}
\begin{eqnarray}
\sigma_\mathrm{v} & \propto & E_z^{1/3}  M_\Delta^{1/3}, \label{eq_evo_2}\\
\lambda & \propto &  M_\Delta , \label{eq_evo_3}\\
L_\mathrm{X} & \propto & E_z^{7/3}  M_\Delta^{4/3}, \label{eq_evo_4}\\
D_\mathrm{A}^2 Y_\Delta & \propto & E_z^{2/3} M_\Delta^{5/3}, \label{eq_evo_5}
\end{eqnarray}
where $D_\mathrm{A}$ is the angular diameter distance to the cluster. The above self-similar scaling relations evolve with redshift as $E_z^{\gamma_\mathrm{ss}}$, with $\gamma_\mathrm{ss}=1/3$, 0, 7/3, or 2/3 for the galaxy velocity dispersion, the optical richness, the X-ray luminosity, or the spherical SZ signal, respectively. The scaling of the X-ray luminosity depends on the energy band. For the soft X-ray luminosity in the rest-frame energy band $[0.1-2.4]~\mathrm{keV}$, $L_\mathrm{X_{soft}}$, the evolution can be expressed as \citep{ett15},
\beq
L_\mathrm{X_{soft}}  \propto  E_z^{2}  M_\Delta .
\eeq

Together with the redshift dependence of the median relation, the intrinsic scatter of the relation and the scatter between the true mass and the mass proxy used to calibrate the relation may evolve as well. Furthermore, any apparent redshift evolution of the scaling may be degenerate with the evolution of either the mass or the selection function. We discuss these aspects in the following.

\subsection{Intrinsic scatter}
\label{sec_scat}

Broadly speaking, the intrinsic scatter of a scaling relation is related to the degree of regularity of the clusters. The larger the deviations from dynamical/hydrostatic equilibrium the larger the scatter \citep{fab+al11,sar+al13}. Scatter is then prominent in morphologically complex halos. Triaxiality is another major source of scatter, since clusters are usually studied under the simplifying assumption of spherical symmetry \citep{lim+al13,ser+al13}. Since high redshift clusters are more irregular and less spherical, the scatter is usually expected to increase with redshift. 

Let us consider the evolution of scatter in a number of cases. Based on numerical simulations, \citet{sar+al13} showed that the scatter of dynamical mass estimates based on the line-of-sight velocity dispersion is approximately log-normal and that it increases with redshift as
\beq
\label{eq_evo_6}
\sigma_{\log(M_{\sigma_\mathrm{v}}/M_\mathrm{Vir})}  \simeq 0.13(1+0.25 z).
\eeq
They argued that the dominant contributor to the scatter is the intrinsic triaxial structure of halos and that its evolution with redshift is also the dominant source of the increasing scatter of the 1D dynamical mass estimates with redshift. 


\citet{fab+al11} studied the scaling relations between the cluster mass and some proxies based on X-ray quantities with a set of cosmological hydrodynamical simulations. They found that the scatter distribution around the best-fitting relations is always close to a log-normal one and that the scatter increases with redshift. 

The precise quantitative estimate of the scatter and of its evolution strongly depends on the details of the baryonic physics included in the simulations. We considered the results of \citet{fab+al11} for runs with non-radiative physics and standard viscosity. To study the time evolution, we fitted the values of the scatter obtained at different redshifts ($z=0.0$, 0.25, 0.50, 0.80, and 1.0) under the assumption of self-similar scaling relation \citep[ tables 1,~2,~and 3]{fab+al11}. The mass proxy $M_{Y_\mathrm{X} }$ is based on $Y_\mathrm{X} $, i.e., the product of the gas mass within $r_{500}$ and the spectroscopic temperature outside the core \citep{kra+al06}. We found that the scatter evolves as
\beq
\label{eq_evo_7}
\sigma_{\log(M_{Y_\mathrm{X} }/M_\mathrm{500})}  \simeq 0.03 E_z^{0.23},
\eeq
or, with an alternative form\footnote{The function $(1+ z)^{\gamma_1}$, i.e., $\sim (1+ \gamma_1 z)$ for $z\ls 1$, can approximate $E_z$ in small redshift intervals. The coefficient $\gamma_1$ used in the approximation depends on the redshift range considered and on the cosmological parameters.},
\beq
\label{eq_evo_8}
\sigma_{\log(M_{Y_\mathrm{X} }/M_\mathrm{500})}  \simeq 0.03(1+0.14 z).
\eeq

For the mass proxy based on the emission-weighted temperature, we found
\beq
\label{eq_evo_9}
\sigma_{\log(M_{T_\mathrm{mw}}/M_\mathrm{500})}  \simeq 0.06 E_z^{0.29},
\eeq
or
\beq
\label{eq_evo_10}
\sigma_{\log(M_{T_\mathrm{mw}}/M_\mathrm{500})}  \simeq 0.06(1+0.19 z).
\eeq

The above results show that the scatter mildly increases with redshift. This suggest that the evolution of the scatter can be modelled as
\beq
\label{eq_evo_11}
\sigma_{o|\mu}  =\sigma_0 E_z^{\gamma_{\sigma}}.
\eeq

\subsection{Completeness}
\label{sec_comp}

The completeness of a sample usually evolves with redshift. Very massive clusters are rare and difficult to be found in the local volume but they are still forming at high redshift. On the other hand, only clusters emitting very strong signals can be detected to very large distances. As detailed in App.~\ref{app_mass}, the selection and the mass functions conjure to make the distribution of true masses in observed samples fairly unimodal. The evolution of the completeness of the sample can be characterised through the evolution of the peak and of the dispersion of this distribution.

The mean (logarithmic) mass of the sample is connected to the observational threshold (see App.~\ref{app_mass}), which may evolve with redshift, and to the scatter between the mass and the observable quantity used to select the clusters, which evolves too.

Let us first consider the evolution of the mass corresponding to a completeness limit. As a first example let us consider a flux-selected sample. The luminosity scales with mass as
\beq
\label{eq_evo_12}
L_\Delta \propto  M_\Delta^\beta E_z^\gamma.
\eeq
If we select only clusters above a limiting flux, $f_\mathrm{th}$, the corresponding luminosity evolves as $L_\mathrm{th}(z) \propto f_\mathrm{th}  D_\mathrm{L}(z)^2$, where $D_\mathrm{L}(z)$ is the luminosity distance. In absence of scatter, the corresponding limiting mass evolves as
\beq
\label{eq_evo_13}
M_\mathrm{th}\propto D_\mathrm{L}(z)^\frac{2}{\beta}E_z^{-\frac{\gamma}{\beta}}.
\eeq

As a second example, let us consider a SZ-like signal, whose size increases with the projected physical surface covered by the cluster. In this case, the observable is proportional to $D_\mathrm{A}(z)^2 \theta_\Delta^2$, where $\theta_\Delta$ is the angular extension of the cluster. The scaling can be written as
\beq
\label{eq_evo_14}
D_\mathrm{A}(z)^2 Y_\Delta \propto M_\Delta^\beta E_z^\gamma .
\eeq
The noise is proportional to the square root of the angular area, i.e., $\sigma_{Y_\Delta} \propto \theta_\Delta = r_\Delta/D_\mathrm{A}$. In absence of scatter, if we select only clusters above a given signal-to-noise ratio (SNR), i.e., $Y_\Delta/\sigma_{Y_\Delta}>\mathrm{SNR_{th}}$, the corresponding threshold mass evolves with redshift as
\beq
\label{eq_evo_15}
M_\mathrm{th}\propto D_\mathrm{A}(z)^\frac{3}{3\beta-1}E_z^\frac{2+3\gamma}{1-3\beta}.
\eeq

The two above examples suggest that the evolution of the mass at a given completeness limit can be parametrized as
\beq
\label{eq_evo_16}
M_\mathrm{th}\propto D_\mathrm{A}(z)^{\gamma_D} E_z^{\gamma_{E_z}},
\eeq
where the factor $E_z^{\gamma_{E_z}}$ accounts for the evolution in both the mass threshold and the intrinsic scatter of the scaling relation. This modelling of the completeness limit was derived for samples selected with a cut on the detection observable but the functional form is flexible enough to address even more complicated cases. The choice of the angular diameter distance over the luminosity distance in Eq.~(\ref{eq_evo_16}) is irrelevant since the two differs for a factor $(1+z)^2$ which can be approximately englobed in $E_z^{\gamma^{E_z}}$ for limited redshift baselines.

The evolution in the dispersion of the mass sample is mainly connected to the intrinsic scatter in the relation used to select the samples, see App.~\ref{app_mass}. The redshift dependence can then be modelled as in Eq.~(\ref{eq_evo_11}).

\section{Regression scheme}
\label{sec_regr}

\begin{table}
\caption{Parameters of the regression scheme and their description. The variables $Z$, $X$, and $Y$ denote (the logarithm of) the true mass, the WL mass, and the self-similarly evolved observable, respectively.}
\label{tab_par}
\centering
\resizebox{\hsize}{!} {
\begin{tabular}[c]{l  l l}
	\hline
	Type & Meaning & Symbol \\ 
	\noalign{\smallskip}  
	\hline
	\multicolumn{3}{l}{$Y_Z = \alpha_{Y|Z}+\beta_{Y|Z} Z + \gamma_z \log E_z$} \\
	\noalign{\smallskip}  
	Conditional scaling relation &	Intercept & $\alpha_{Y|Z}$ \\
	 & Mass evolution & $\beta_{Y|Z}$ \\
	& Time evolution & $\gamma_z$ \\
	\noalign{\smallskip}
	\multicolumn{3}{l}{$\sigma_{Y|Z}(z)=\sigma_{Y|Z,0} E_z^{\gamma_{\sigma_{Y|Z}}}$} \\  
	 \noalign{\smallskip}  
	 Conditional intrinsic scatter & Conditional scatter at $z=0$	& $\sigma_{Y|Z,0}$ \\
	 &	Time evolution				& $\gamma_{\sigma_{Y|Z}}$ \\
	\noalign{\smallskip}  
	\multicolumn{3}{l}{$\sigma_{X|Z}(z)=\sigma_{X|Z,0} E_z^{\gamma_{\sigma_{X|Z}}}$} \\  
	 \noalign{\smallskip}  
	 Intrinsic scatter of the WL mass  & Conditional scatter at $z=0$	& $\sigma_{X|Z,0}$ \\
	 & Time evolution				& $\gamma_{\sigma_{X|Z}}$ \\
	 \noalign{\smallskip}  
	\multicolumn{3}{l}{$\mu_Z (z) = \bar{\mu}_Z +\gamma_{\mu_Z}\log E_z + \gamma_{\mu_Z,D}\log D_\mathrm{A}(z)$} \\ 
	 \noalign{\smallskip}  
	 Mean of the mass function &  Normalisation & $\bar{\mu}_Z$ \\
	 & Time evolution with $E_z$& $\gamma_{\mu_Z}$ \\
	& Time evolution with $D_\mathrm{A}$ & $\gamma_{\mu_Z,D}$ \\
	 \noalign{\smallskip}  
	\multicolumn{3}{l}{ $\sigma_{Z}(z)=\sigma_{Z,0}E_z^{\gamma_{\sigma_{Z}}}$} \\
        \noalign{\smallskip}  
        Dispersion of the mass function & Dispersion at $z=0$	& $\sigma_{Z,0}$ \\
       & Time evolution	& $\gamma_{\sigma_{Z}}$ \\
	\hline
	\end{tabular}
	}
\end{table}

When we calibrate a scaling relation we deal with: {\it i}) the true mass of the cluster $M_\Delta$, which we cannot measure; {\it ii}) a scattered (and likely biased) proxy of the true mass, such as the weak lensing mass $M_{\mathrm{WL},\Delta}$, which is the proxy we considered in following, or the hydrostatic mass $M_{\mathrm{HE},\Delta}$ (see \S~2 and app.~A of \citetalias{se+et14_comalit_I}); {\it iii}) an observable quantity $O$, which we assume to be on average related to the true mass with a power-law.
 
In logarithmic variables, the median scaling relation is approximatively linear and the scatter is Gaussian. As discussed in Sec.~\ref{sec_evol}, the scaling can be expressed as 
\beq
\label{eq_bug_0}
\log (E_z^{-\gamma_\mathrm{ss}} O) = \alpha+ \beta \log M_\Delta+ \gamma_z \log E_z.
\eeq
Since we englobed the self-similar evolution in the left-hand of Eq.~(\ref{eq_bug_0}), values of the parameters $\gamma_z$ which are different from zero denote deviations from the self-similar time dependence. In other words, the time evolution of the scaling relations $\gamma_z $ is relative to that predicted by the self-similar model. Given a particular scaling law, there is negative, i.e., $\gamma_z<0$ (positive, i.e., $\gamma_z>0$) evolution if the normalisation at high redshift is lower (higher) than anticipated from the self-similar scaling.

In what follows, which is the general scheme we employed for the regression analysis, we identify $\log M_\Delta$ with the variable $Z$, we identify the logarithm of the mass proxy, i.e., the weak lensing mass, with the random variable $X$, and we identify the logarithm of the self-similarly redshift evolved observable with the response $Y$. In this scheme, the mass is the covariate variable, as when using number counts of galaxy clusters to constrain cosmological parameters. The observed values are denoted with the lower case, i.e., $x$ and $y$ are the manifest measured estimates of the latent $X$ and $Y$, respectively \citep{fe+ba12}. This notation is the convention adopted in the CoMaLit series.

The conditional probability of $X$ given $Z$ is
\beq
\label{eq_bug_1}
P(X|Z) = {\cal N}(Z, \sigma_{X|Z}(z)),
\eeq
where ${\cal N}$ is the normal distribution. In Eq.~(\ref{eq_bug_1}), $X$ is an unbiased proxy of $Z$. Any bias between $X$ and $Z$ would be degenerate with the estimated overall normalisation of the scaling between $Y$ and $Z$. This bias cannot be determined with the data, which only constrain the relative bias between $X$ and $Y$ \citepalias[see][]{se+et14_comalit_I}. As discussed in Section~\ref{sec_scat}, the redshift evolution of the scatter is modelled as
\beq
\label{eq_bug_2}
\sigma_{X|Z}(z)=\sigma_{X|Z,0}E_z^{\gamma_{\sigma_{X|Z}}}.
\eeq
The mean observable for a given mass is linearly related to the (logarithm of the) mass and the relation evolves with redshift,
\beq
\label{eq_bug_3}
Y_Z = \alpha_{Y|Z}+\beta_{Y|Z} Z + \gamma_z \log E_z;
\eeq
the redshift $z$ is deterministic and assumed to be known without measurement errors. $Y$ is scattered and distributed according to the conditional probability 
\beq
\label{eq_bug_4}
P(Y|Z) = {\cal N}(Y_Z, \sigma_{Y|Z}(z)),
\eeq
with
\beq
\label{eq_bug_5}
\sigma_{Y|Z}(z)=\sigma_{Y|Z,0} E_z^{\gamma_{\sigma_{Y|Z}}}.
\eeq
The distribution of the masses can be approximated with a Gaussian function
\beq
\label{eq_bug_6}
P(Z) = {\cal N}(\mu_Z (z), \sigma_{Z}(z)).
\eeq
The mass distribution resulting from usual selection procedures is fairly unimodal (see App.~\ref{app_mass}) and can be approximated with the normal distribution of Eq.~(\ref{eq_bug_6}). The statistical improvement obtained considering more complex distributions, such as mixture of Gaussians with different means and variances, is usually negligible (\citealt{kel07};\citetalias{ser+al14_comalit_II}).

As discussed in Sec.~\ref{sec_comp}, the evolution of the (mean of the) mass function can be modelled after Eq.~(\ref{eq_evo_16}) as
\beq
\label{eq_bug_7}
\mu_Z (z) = \bar{\mu}_Z +\gamma_{\mu_Z}\log E_z + \gamma_{\mu_Z,D}\log D_\mathrm{A}(z).
\eeq
The dispersion evolves as
\beq
\label{eq_bug_8}
\sigma_{Z}(z)=\sigma_{Z,0}E_z^{\gamma_{\sigma_{Z}}}.
\eeq

The completeness function at a given redshift can be computed by dividing the estimated mass function (Eq.~\ref{eq_bug_6}) by the cosmological halo mass function. This approach requires the knowledge of the effective survey area of the sample, which may be difficult to estimate for heterogeneous samples. Alternatively, we can use the approximate formulae presented in App.~\ref{app_mass}, which were derived under the assumptions that the completeness function can be approximated as a complementary error function and that the cosmological halo mass function can be approximated as a power-law.

If we assume that the uncertainty in the measurement process is Gaussian, the relation between the unknown $X_i$ and $Y_i$ and the measured $x_i$ and $y_i$ is given by
\beq
\label{eq_bug_9}
P(\{x_i,y_i\}|\{X_i,Y_i\}) = {\cal N}^\mathrm{2D}(\{X_i,Y_i\}, \mathbfss{V}_i) {\cal U}(y_\mathrm{th,i},\infty),
\eeq
where ${\cal N}^\mathrm{2D}$ and ${\cal U}$ are the bivariate Gaussian and the uniform distribution, respectively. In Eq.~(\ref{eq_bug_9}), $\mathbfss{V}_i$ is the symmetric uncertainty covariance matrix whose diagonal elements are denoted as $\delta_{x,i}^2$ and $\delta_{y,i}^2$, and whose off-diagonal elements are denoted as $\rho_{xy}\delta_{x,i}\delta_{y,i}$.

The truncation, i.e., null probability for $y_i<y_\mathrm{th,i}$, accounts for selection effects when only clusters above an observational limit (in the response variable) are included in the sample, i.e., the Malmquist bias \citepalias{ser+al14_comalit_II}.

The treatment is complete once the priors on the parameters are made explicit. We choose non-informative priors as discussed in \citetalias{se+et14_comalit_I} and \citetalias{ser+al14_comalit_II}. The priors on the intercept $\alpha_{Y|Z}$ and on the mean $\bar{\mu}_Z$ are taken to be flat,
\beq
\label{eq_bug_10}
\alpha_{Y|Z},\  \bar{\mu}_Z  \sim  {\cal U}(-1/\epsilon,1/\epsilon),
\eeq
where $\epsilon$ is a small number. In our calculation we took $\epsilon = 10^{-3}$. A priori, the slopes follow the Student's $t_1$ distribution with one degree of freedom, as suitable for uniformly distributed direction angles,
\beq
\label{eq_bug_11}
\beta_{Y|Z},\ \gamma_z,\ \gamma_{\sigma_{X|Z}},\ \gamma_{\sigma_{Y|Z}},\ \gamma_{\mu_Z},\ \gamma_{\mu_Z,D},\ \gamma_{\sigma_Z} \sim  t_1.
\eeq
The Student prior for the slopes is not informative. Negative time evolutions and scatters which decreases at early times are allowed. For the variances, we adopted an inverse Gamma distribution,
\beq
\label{eq_bug_12}
1/\sigma_{X|Z,0}^2,\ 1/\sigma_{Y|Z,0}^2,\ 1/\sigma_{Z,0}^2 \sim \Gamma(\epsilon,\epsilon).
\eeq

This regression scheme requires 12 parameters, i.e., three parameters characterising the scaling relation, two for the intrinsic scatter, two for the mass scatter, and five for the mass function, plus three variables for each cluster, i.e., the true weak lensing mass, the true mass, and the true observable. The parameters and their meanings are summarised in Table~\ref{tab_par}.

The relation in Eq.~(\ref{eq_bug_3}) expresses the conditional scaling relation, wherein $Y_Z$ is the most likely value of the variable $Y$ for a given $Z$. This is the relation to be used to predict the value of $Y$ for a given $Z$. The relation between two random and scattered variables might be better described by the symmetric scaling relation, which goes along the direction where the probabilities of $Z$ and $Y$ are maximised at the same time \citepalias{ser+al14_comalit_II}. In the above regression scheme, where $Z$ and $Y$ follow a bivariate normal distribution, the slope of the symmetric scaling can be expressed as \citepalias{ser+al14_comalit_II}
\beq
\label{eq_slop_8}
\beta_{Y\textrm{-}Z} =\frac{\sigma_Z^2-\sigma_Y^2-\sqrt{\sigma_Z^4+2(2\rho_{YZ}^2-1)\sigma_Z^2\sigma_Y^2+\sigma_Y^4}}{2\rho_{YZ}\ \sigma_Z \sigma_Y},
\eeq
where $\rho_{YZ}$ is the correlation factor between $Y$ and $Z$, 
\beq
\rho_{YZ}=\frac{\beta_{Y|Z}}{\sqrt{\sigma_{Y|Z}^2/\sigma_Z^2-\beta_{Y|Z}^2}},
\eeq
and the variance in $Y$ is related to the conditional scatter as
\beq
\sigma_Y^2=\sigma_{Y|Z}^2+\beta_{Y|Z}^2\sigma_Z^2.
\eeq
The intercept of the symmetric relation can be expressed as
\beq
\alpha_{Y\textrm{-}Z}=\alpha_{Y|Z}+(\beta_{Y|Z}-\beta_{Y\textrm{-}Z}) \mu_Z.
\eeq

The detailed regression scheme is simplified when we are interested in the scaling between the observable and the measured proxy mass, i.e., the WL or the X-ray mass,
\beq
\label{eq_bug_13}
\log (E_z^{-\gamma_\mathrm{ss}} O) = \alpha+ \beta \log M_\mathrm{proxy}+ \gamma_z \log E_z.
\eeq
In this case, the adopted form for the scaling is
\beq
\label{eq_bug_14}
Y_X=\alpha_{Y|X}+\beta_{Y|X}X+\gamma_z\log E_z,
\eeq
which substitutes Eq.~(\ref{eq_bug_3}). The latent variable $Z$ coincides now with the manifest one $X$ and we do not have to model the conditional probability of $X$ given $Z$, see Eqs.~(\ref{eq_bug_1}) and (\ref{eq_bug_2}).

\section{Cluster catalogs}
\label{sec_cata}

There are different approaches to choose a sample of clusters to analyse. We may look for a statistical sample which is complete with respect to well-defined selection criteria. This sample would be ideal but most of the massive clusters with very good quality data might be excised. The alternative is to assemble samples as numerous as possible with the idea that variety and largeness can compensate for incompleteness and inhomogeneity.

These two approaches are to some degree complementary and have been already discussed in \citetalias{ser+al14_comalit_II} and \citetalias{ser14_comalit_III}, which we refer to for further considerations. Here we are mainly interested in testing the regression algorithm and we focus on large samples. To this aim we assembled a catalog of clusters with measured velocity dispersions. As catalogs of weak lensing masses, optical richnesses, X-ray luminosities, and SZ effects, we used publicly available compilations. The subsample of weak lensing clusters also included in either the velocity dispersion, richness, X-ray, or SZ catalogs were used in the analysis presented in the next section. We briefly discuss the main properties of the catalogs and refer to the original references for further details.

\subsection{Weak lensing masses}

\citetalias{ser14_comalit_III} retrieved from literature 822 weak lensing analyses of clusters and groups with measured redshift and mass. Here, we consider the LC$^2$-{\it single}, which contains 485 unique entries with reported coordinates, redshift, and WL masses to over-densities of 2500, 500, 200, and to the virial radius.\footnote{The catalogs are available at \url{http://pico.bo.astro.it/\textasciitilde sereno/CoMaLit/LC2/}} Duplicate entries from input references were carefully handled.

The cluster redshifts span a large interval, $0.02\ls z\ls 1.5$. The catalog is large and standardised but it is not statistically complete. We refer to \citetalias{ser14_comalit_III} for a detailed discussion of the catalog properties.

\subsection{Velocity dispersions}

We assembled some publicly available catalogs of clusters with measured velocity dispersions. We first review the source catalogs and then we introduce the merged compilation.

\subsubsection{Source catalogs}

\citet{cav+al09} presented the results from the spectroscopic survey WINGS (WIde-field Nearby Galaxy-cluster Survey)-SPE, which consists of 48 nearby clusters at $0.04\ls z\ls 0.07$ selected from three X-ray flux limited samples. They complemented the sample with 29 additional clusters not observed in the programme but for which literature data existed. The total sample contains 77 clusters over a broad range of richness, Bautz-Morgan class, and X-ray luminosity.

\citet{ebe+al07} presented the sample of the 12 most distant galaxy clusters detected at $z\gs0.5$ by the Massive Cluster Survey (MACS). This catalog is statistically complete and comprehensive of measurements of radial velocity dispersions.

\citet{gi+me01} considered a sample of 51 distant galaxy clusters at $0.15\ls z\ls0.9$, each cluster having at least 10 galaxies with available redshift in the literature. In some clusters, two peaks that are not clearly separable were identified in the velocity distribution. For these systems with uncertain internal dynamics, we considered the velocity dispersion measured by analysing the identified peaks together. We also discarded two systems with no major peak (CL~J0023+0423 and CL~J0949+44).

\citet{maz+al96} constructed a volume limited sample of 128 clusters out to $z=0.1$ combining data from the ENACS (ESO Nearby Abell Clusters Survey) with pre-existing data from literature. They measured reliable velocity dispersions for a subset of 80 of them, based on at least 10 redshifts. They also analysed 26 additional clusters in the cone but with $z>0.1$. The total catalog consists of 106 clusters. We discarded from our final catalog the secondary systems.

\citet{oe+hi01} presented the spectroscopic study of a sample of 25 Abell clusters out to $z=0.1$ containing a central cD galaxy. Redshifts measured with the MX Spectrometer were combined with those collected from the literature to obtain typically 50--150 observed velocities in each cluster. We used the estimates of the velocity dispersions within the smaller quoted aperture ($\sim 1.2~\mathrm{Mpc}/h$ at $z=0.1$)

\citet{pop+al07} considered a sample of 137 optically selected and spectroscopically confirmed Abell clusters in the SDSS (Sloan Digital Sky Survey) database \citep{ade+al06}. The clusters span the redshift range $0.04 \ls z \ls 0.17$. 40 of the clusters were X-ray under-luminous, since they had a marginal X-ray detection or remained undetected in the ROSAT All Sky Survey.

\citet{ri+di06} studied the infall patterns of 72 nearby ($z<0.1$) clusters from the Data Release (DR) 4 of the SDSS. The clusters were selected in X-ray flux from the ROSAT All-Sky Survey. Velocity dispersions were measured and masses were derived with the caustic method. \citet{ri+di10} extended the approach to a sample of 16 groups with lower X-ray fluxes selected from the 400 deg$^2$ serendipitous survey of clusters. Spectroscopic data were taken from the SDSS DR5. \citet{rin+al13} selected 58 clusters by their X-ray flux and in the redshift interval $0.1<z<0.3$ to build the Hectospec Cluster Survey (HeCS), the first systematic spectroscopic survey of cluster infall regions at $z \ga 0.1$. For each cluster, high signal-to-noise spectra for $\sim$200 cluster members were acquired with MMT (Multi Mirror Telescope)/Hectospec.

\citet{rue+al14} presented optical spectroscopy of galaxies in clusters detected through the SZ effect with the South Pole Telescope (SPT). They reported measurements of 61 spectroscopic cluster redshifts, and 48 velocity dispersions each calculated with more than 15 member galaxies. After the inclusion of additional measurements of SPT-observed clusters previously reported in the literature, the final catalog presents 57 velocity dispersions. Being SZ selected, most of the clusters are at high redshift. The clusters span an interval $0.3 \ls z \ls 1.5$.

\citet{sif+al13} presented the dynamical analysis of a sample of 16 SZ selected massive clusters detected with the Atacama Cosmology Telescope (ACT)  over a 455 $\deg^2$ area of the southern sky. 60 member galaxies on average per cluster were observed with deep multi-object spectroscopic observations. The sample spans the redshift range $0.3 \ls z \ls 1.1$ with a median redshift $z=0.50$.

\citet{zha+al11} presented a multi-wavelength analysis of 62 galaxy clusters in the HIFLUGCS (HIghest X-ray FLUx Galaxy Cluster Sample), an X-ray flux-limited sample. Velocity dispersions were computed thanks to 13439 cluster member galaxies with redshifts collected from literature. Most of the clusters (60 out of 62) are at $z <0.1$.

\subsubsection{Merged catalog}

The catalogs listed before provides a total of 710 velocity dispersion estimates, comprehensive of multiple peaks and substructures which we did not consider in our final sample.

Cluster coordinates were taken from the original or from companion papers. When they were not reported, we used the coordinates listed by the NASA/IPAC Extragalactic Database (NED).\footnote{\url{http://ned.ipac.caltech.edu/}.}

Not unique entries were identified by matching names and cluster coordinates. For clusters with multiple analysis, we preferred the study based on the larger number of identified cluster member galaxies with measured redshift, $N_\mathrm{members}$. The final catalog contains 564 unique clusters. The catalogs are publicly available at \url{http://pico.bo.astro.it/\textasciitilde sereno/CoMaLit/sigma/}. Their format is detailed in App.~\ref{app_SC}.

When original estimates were provided with asymmetric errors, we computed the mean value as suggested in \citet{dag04}. To standardise the uncertainties, we followed \citet{rue+al14}. They found that the uncertainty in the velocity dispersion $\sigma_\mathrm{v}$ is well described by
\beq
\delta_{\sigma_\mathrm{v}}= \frac{0.92\sigma_\mathrm{v}}{\sqrt{N_\mathrm{members}-1}}
\eeq
when including the effect of membership selection.

As we inferred from the matching with the LC$^2$-{\it single}, WL masses are known for a subsample of 97 clusters. This size can be achieved only relying on a number of different source catalogs. 30 clusters are from \citet{gi+me01}, 23 from \citet{rin+al13}, 13 from \citet{zha+al11}, 11 from \citet{ebe+al07}, 8 from \citet{rue+al14}, 5 from \citet{pop+al07}, 4 from \citet{maz+al96}, and 3 from \citet{sif+al13}.

\subsection{Optical richness}
\label{sec_cata_redmapper}

\citet{ryk+al14} applied the redMaPPer (red-sequence Matched-filter Probabilistic Percolation), a red-sequence cluster finder designed for large photometric surveys, to $\sim10000~\deg^2$ of SDSS DR8 data. The resulting catalog\footnote{We used the latest version of the catalog (v5.2), which is publicly available at \url{http://risa.stanford.edu/redMaPPer/}.} contains $\sim 25000$ candidate clusters over the redshift range $0.08 \ls z \ls 0.55$. 

According to the catalog convention, the richness $\lambda$ of a cluster is defined as the sum of the probabilities of the galaxies found near a cluster to be actually cluster members. The sum extends over all galaxies above a cut-off luminosity ($0.2L_*$) and below a radial cut which scales with richness. Clusters are included in the catalog if their richness exceeds 20 times the scale factor $S_\mathrm{RM}$ (also provided in the catalog), which is a function of the photometric redshift of the cluster. This selection criterion approximately requires that every cluster has at least 20 galaxy counts above the flux limit of the survey or the cut-off luminosity at the cluster redshift, whichever is higher.

\subsection{X-ray clusters}

\citet{mau+al08} presented a sample of 114 clusters covering wide temperature ($2 \la k_\mathrm{B}T \la16~\mathrm{keV}$) and redshift ($0.1 \la z \la 1.3$) baselines. The sample was assembled from all publicly available {\it Chandra} data as of 2006 November. It consists of clusters at redshift greater than 0.1 listed in the NED which were the targets of observations made with the ACIS-I detector covering at least half of the area in the annulus [0.9--1.0]~$r_{500}$.  The radius $r_{500}$ was estimated assuming a $M_{500}$-$Y_\mathrm{X} $ relation. 

The sample was later reanalysed using updated softwares and calibration files in \citet{mau+al12}. Bolometric luminosities were measured either in the [0-1]~$r_{500}$ aperture, which we took as reference case to ease the comparison with theoretical predictions, or in the core-excised [0.15--1]~$r_{500}$ aperture, $L_\mathrm{X,ce}$. This large catalog is standardised in the measurement procedures but it is not statistically complete.

As an alternative we also looked at catalogs of X-ray luminosities measured in the [0.1--2.4] keV band, $L_\mathrm{X_{soft}}$. We considered the MCXC \citep[Meta-Catalogue of X-ray detected Clusters of galaxies,][]{pif+al11}, which comprises 1743 unique X-ray clusters collected from available ROSAT All Sky Survey-based and serendipitous cluster catalogues. X-ray luminosities were systematically homogenised and standardised to an over-density of $\Delta=500$. Uncertainties are not provided in the catalog. For our tests, we fixed the statistical uncertainty to 10 per cent. As the LC$^2$, the MCXC is not statistically complete.

\subsection{Planck SZ catalog}

The {\it Planck} SZ Catalogue \citep[PSZ,][]{planck_2013_XXIX} contains 883 candidates identified with the Matched Multi-filter method MMF3 with detections above SNR = 4.5. The catalog spans a broad mass range from $0.1$ to $16\times10^{14}M_\odot$ at a median redshift of $z\sim 0.22$. The redshift determination is available for 664 candidates.

In \citetalias{ser+al14_comalit_II}, we computed the spherically integrated $Y_{500}$ of the PSZ clusters within the weak-lensing determined $r_{500}$. The measurements of $M_{\mathrm{WL},500}$ and $Y_{500}$ are then correlated. In our analysis, we used the full uncertainty covariance matrix. We refer to \citetalias{ser+al14_comalit_II} for a detailed discussion.

\section{Results}
\label{sec_resu}

\begin{figure*}
\resizebox{\hsize}{!} {
 \includegraphics{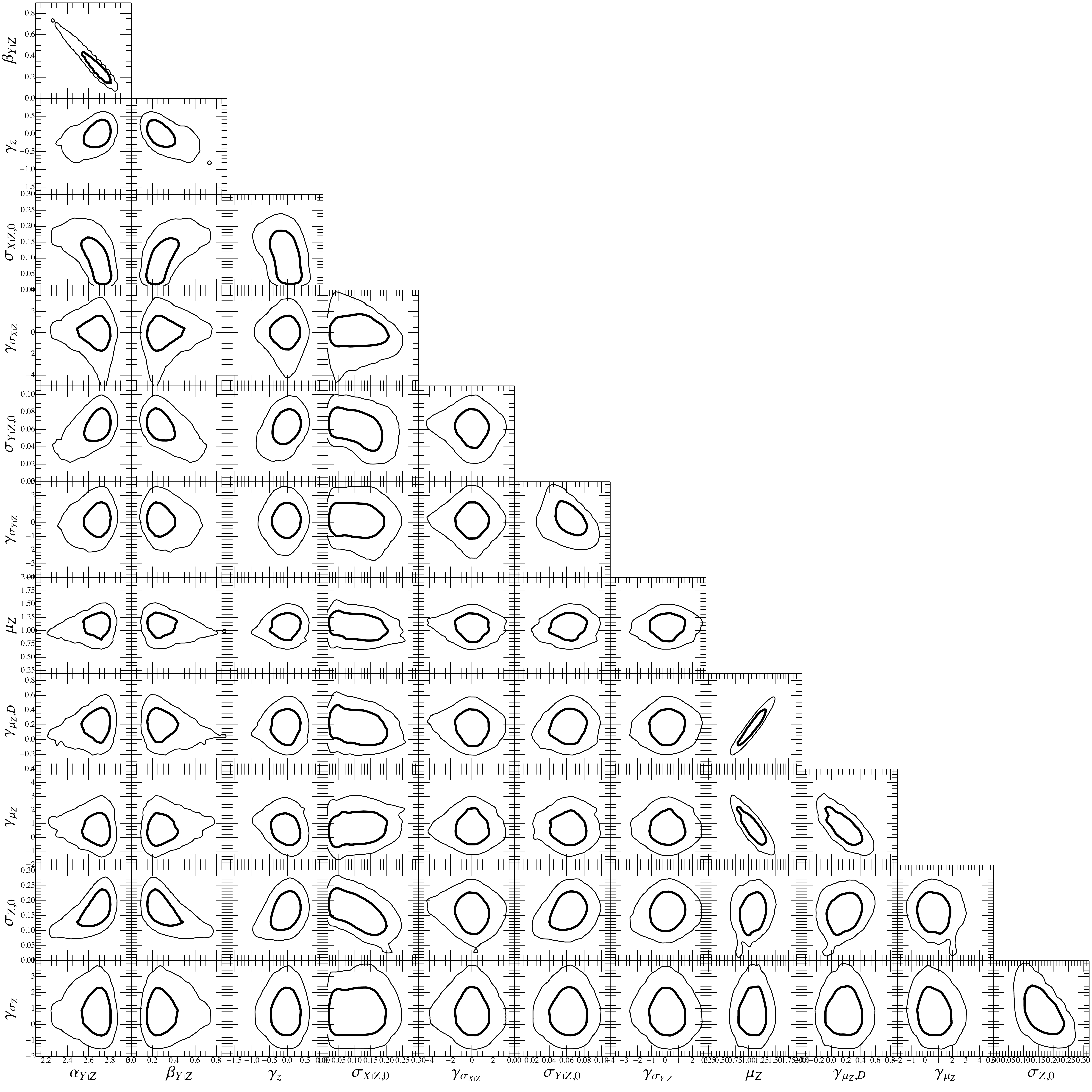} }
\caption{Probability distributions of parameters of the scaling relation between velocity dispersion and mass, $\sigma_\mathrm{v}$-$M_{200}$, and of the mass function. The thick (thin) lines include the 1-(2-)$\sigma$ confidence region in two dimensions, here defined as the region within which the value of the probability is larger than $\exp[-2.3/2]$ ($\exp[-6.17/2]$) of the maximum.}
\label{fig_sigma_M_PDF}
\end{figure*}

\begin{figure*}
\resizebox{\hsize}{!} {
 \includegraphics{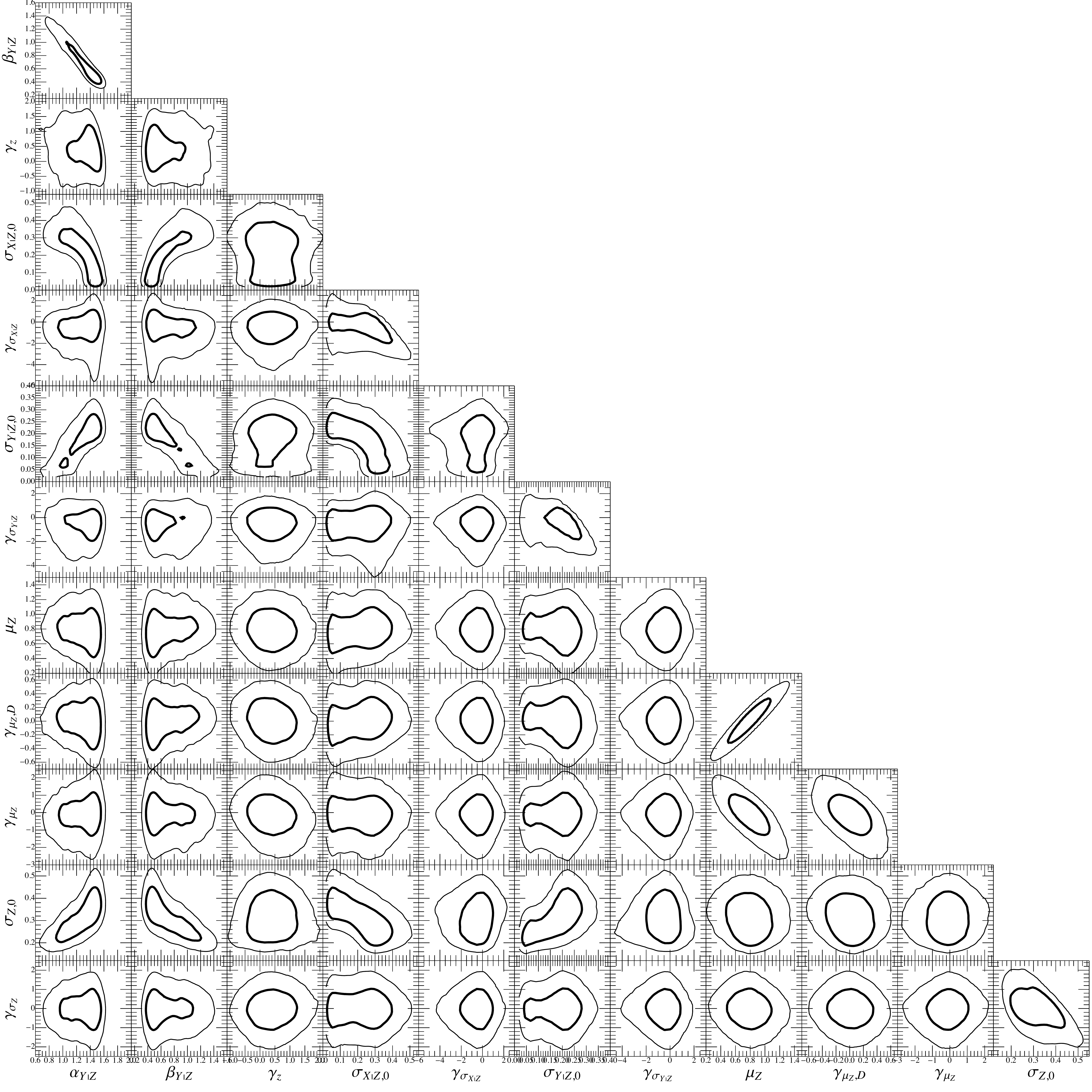} }
\caption{Probability distributions of parameters of the scaling relation between optical richness and mass, $\lambda$-$M_{200}$, and of the mass function. The thick (thin) lines include the 1-(2-)$\sigma$ confidence region in two dimensions, here defined as the region within which the value of the probability is larger than $\exp[-2.3/2]$ ($\exp[-6.17/2]$) of the maximum.}
\label{fig_richness_M_PDF}
\end{figure*}

\begin{figure*}
\resizebox{\hsize}{!} {
 \includegraphics{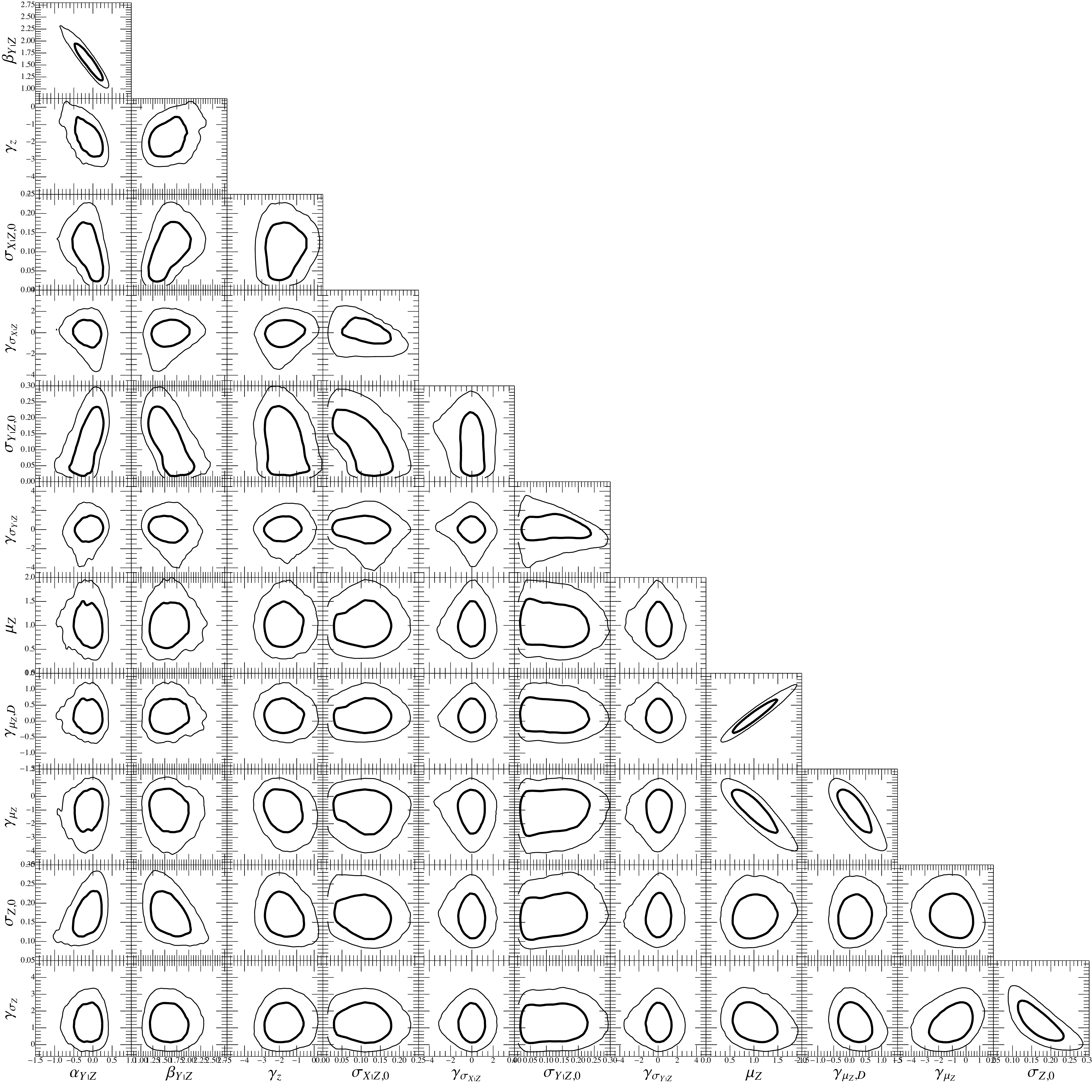} }
\caption{Probability distributions of parameters of the scaling relation between X-ray luminosity and mass, $L_\mathrm{X}$-$M_{500}$, and of the mass function. The thick (thin) lines include the 1-(2-)$\sigma$ confidence region in two dimensions, here defined as the region within which the value of the probability is larger than $\exp[-2.3/2)]$ ($\exp[-6.17/2]$) of the maximum.}
\label{fig_LX_M_PDF}
\end{figure*}

\begin{figure*}
\resizebox{\hsize}{!} {
 \includegraphics{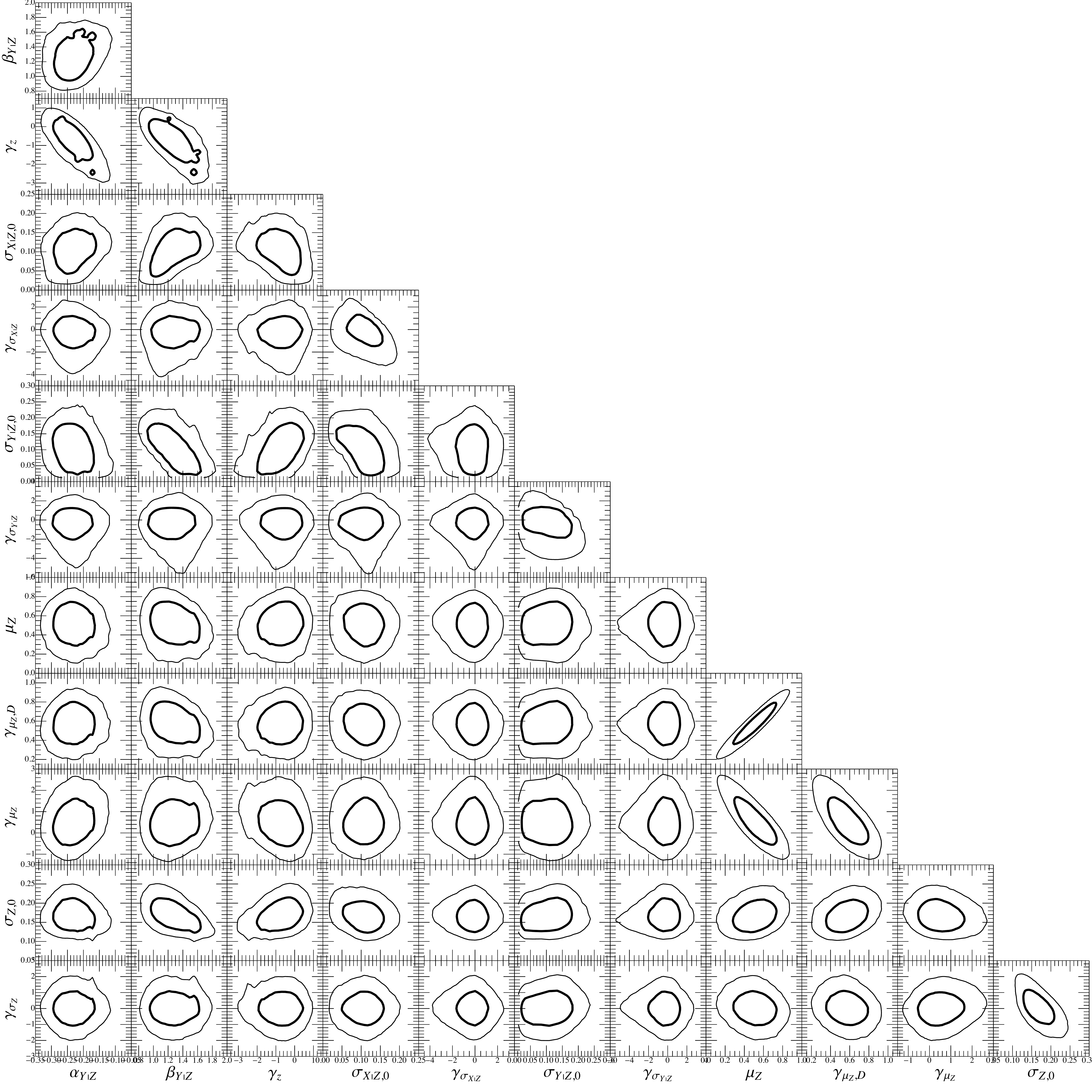} }
\caption{Probability distributions of parameters of the scaling relation $Y_\mathrm{SZ}$-$M_{500}$, and of the mass function. The thick (thin) lines include the 1-(2-)$\sigma$ confidence region in two dimensions, here defined as the region within which the value of the probability is larger than $\exp[-2.3/2]$ ($\exp[-6.17/2]$) of the maximum.}
\label{fig_YSZ_M_PDF}
\end{figure*}

\begin{table*}
\caption{Observed scaling relations, $E_z^{-\gamma_\mathrm{ss}}O=10^{\alpha} M_\Delta^{\beta}E_z^{\gamma_z}$. Conventions are as in Section~\ref{sec_regr} and Table~\ref{tab_par}: $Z=\log (M_{\Delta})$; $X=\log (M_{\mathrm{WL},\Delta})$; $Y=\log(E_z^{-\gamma_\mathrm{ss}}\sigma_\mathrm{v})$, $\log(E_z^{-\gamma_\mathrm{ss}}\lambda)$, $\log(E_z^{-\gamma_\mathrm{ss}}L_\mathrm{X})$, or $\log (E_z^{-\gamma_\mathrm{ss}}D_\mathrm{A}^{2} Y_\mathrm{500})$. $\gamma_\mathrm{ss}$ is the self-similar evolution, which is equal to 1/3, 0, 7/3, 2/3 for $\sigma_\mathrm{v}$--, $\lambda$--, $L_\mathrm{X}$--, and $Y_\mathrm{SZ}$--$M_\Delta$, respectively. Units are $10^{14}M_\odot$ for mass, km/s for $\sigma_\mathrm{v}$,  $10^{44}$~ergs~s$^{-1}$ for the bolometric luminosity $L_\mathrm{X}$, and $10^{-4}$Mpc$^2$ for $D_\mathrm{A}^2 Y_\mathrm{500}$. Cols.~1--3: variables of the regression procedure. Col.~4: number of clusters in the sample ($N_\mathrm{cl}$). Col.~5: median redshift of the sample. Cols.~6, 7, and 8: intercept, mass slope, and time evolution of the conditional scaling relation. Cols.~9-10: local scatter of the WL mass and its time-evolution. Cols.~11-12: intrinsic scatter of the scaling relation and its time-evolution. Cols.~13-14:  intercept and slope of the symmetric scaling relation at the median redshift of the sample. Col.~15: self-similar prediction for the slope ($\beta_\mathrm{ss}$).
}
\label{tab_scaling}
\centering
\resizebox{\hsize}{!} {
\begin{tabular}[c]{l  l  l r r r@{$\,\pm\,$}l  r@{$\,\pm\,$}l  r@{$\,\pm\,$}l  r@{$\,\pm\,$}l  r@{$\,\pm\,$}l  r@{$\,\pm\,$}l  r@{$\,\pm\,$}l  r@{$\,\pm\,$}l  r@{$\,\pm\,$}l l}
	\hline
	\multicolumn{5}{c}{} & \multicolumn{6}{c}{Conditional scaling} & \multicolumn{4}{c}{WL mass scatter}  & \multicolumn{4}{c}{Intrinsic scatter} & \multicolumn{4}{c}{Symmetric scaling} & \\ 
	\noalign{\smallskip}  
	$10^Z$ & $10^X$ &  $10^Y$& 	$N_\mathrm{Cl}$& $z$&	\multicolumn{2}{c}{$\alpha_{Y|Z}$}	&	\multicolumn{2}{c}{$\beta_{Y|Z}$}&	\multicolumn{2}{c}{$\gamma_z$}&	\multicolumn{2}{c}{$\sigma_{X|Z,0}$}&	\multicolumn{2}{c}{$\gamma_{\sigma_{X|Z}}$}&	\multicolumn{2}{c}{$\sigma_{Y|Z,0}$}&	\multicolumn{2}{c}{$\gamma_{\sigma_{Y|Z}}$}  & \multicolumn{2}{c}{$\alpha_{Y\textrm{-}Z}$}	& \multicolumn{2}{c}{$\beta_{Y\textrm{-}Z}$}	&	$\beta_\mathrm{ss}$\\
	\hline
	$M_{200}$	&	$M_{\mathrm{WL},200}$	&		$E_z^{-1/3}\sigma_\mathrm{v}$	 & 97&	0.23&	
	2.67&0.12	&   0.30&0.13&    -0.05&0.30&    0.11&0.06&    0.00&1.26&    0.06&0.02&    0.17&0.96   &
	2.63&0.14& 	0.34&0.15&	1/3\\
	$M_{200}$	&	$M_{\mathrm{WL},200}$	&	$\lambda$&	 157& 	0.30&	
	1.29&0.21&   0.70&0.27&   0.39&0.58&   0.22&0.12&   -0.56&1.22&   0.18&0.08&   -0.55&1.11&
	1.12&0.21&	0.91&0.25&	1 \\
	$M_{500}$	&	$M_{\mathrm{WL},500}$	&	$E_z^{-7/3}L_\mathrm{X}$ & 73&	0.38&	
	-0.13&0.26&    1.60&0.27&    -1.74&0.75&    0.11&0.05&    -0.04&1.01&    0.12&0.07&    0.03&1.16 &
	-0.29&	0.24&	1.78&	0.25& 	4/3\\
	$M_{500}$	&	$M_{\mathrm{WL},500}$	&	$E_z^{-2/3}D_\mathrm{A}^2 Y_{500}$ & 115&	0.23&	
	-0.23&0.04&    1.27& 0.21&    -0.76& 0.80&   0.11&0.04&    -0.25& 1.20&    0.11&0.05&    -0.37&1.32 &
	-0.27&0.06&	1.50&0.21&	5/3\\
	\hline
	\end{tabular}
	}
\end{table*}

\begin{table*}
\caption{Mass functions of the observed samples. Conventions and units are as in Tables~\ref{tab_par} and~\ref{tab_scaling}.}
\label{tab_mass_func}
\centering
\begin{tabular}[c]{l  r@{$\,\pm\,$}l  r@{$\,\pm\,$}l  r@{$\,\pm\,$}l  r@{$\,\pm\,$}l  r@{$\,\pm\,$}l }
	\hline
	 & \multicolumn{6}{c}{Mean} & \multicolumn{4}{c}{Dispersion} \\ 
	\noalign{\smallskip}  
	Sample &	\multicolumn{2}{c}{$\bar{\mu}_Z$}&	\multicolumn{2}{c}{$\gamma_{\mu_Z,D}$}&	\multicolumn{2}{c}{$\gamma_{\mu_Z}$}&	\multicolumn{2}{c}{$\sigma_{Z,0}$}&	\multicolumn{2}{c}{$\gamma_{\sigma_{Z}}$}  \\
	\hline
	$\sigma_\mathrm{v}$-$M_{200}$ &	1.08&0.17&    0.18&0.16&    0.69&0.86&    0.16&0.04&    0.74&0.96 \\
	$\lambda$-$M_{200}$&    		0.79& 0.21&   0.01& 0.24&   -0.13&0.92&    0.31&0.07&    -0.04& 0.77 \\
	$L_\mathrm{X}$-$M_{500}$ &				1.02&0.32&    0.17&0.36&    -1.03&1.08&   0.17&0.04&    1.30& 0.77 \\
	$Y_\mathrm{SZ}$-$M_{500}$	&	0.51&0.15&     0.57&0.15&    0.53&0.79&     0.17&0.03&     0.03&0.79 \\
	\hline
	\end{tabular}
\end{table*}

\begin{table*}
\caption{Scaling relations as a function of the WL mass, $E_z^{-\gamma_\mathrm{ss}}O=10^{\alpha} M_{\mathrm{WL},\Delta}^{\beta}E_z^{\gamma_z}$. Listed parameters refer to logarithmic variables. For the conditional scaling, the adopted form is $Y_X=\alpha_{Y|X}+\beta_{Y|X}X+\gamma_z\log E_z$. Conventions and units are as in Sec.~\ref{sec_regr} and Tables~\ref{tab_par} and~\ref{tab_scaling}.}
\label{tab_scaling_MWL}
\centering
\begin{tabular}[c]{l   r@{$\,\pm\,$}l  r@{$\,\pm\,$}l  r@{$\,\pm\,$}l  r@{$\,\pm\,$}l  r@{$\,\pm\,$}l r@{$\,\pm\,$}l r@{$\,\pm\,$}l}
	\hline
	 & \multicolumn{6}{c}{Conditional scaling}  & \multicolumn{4}{c}{Intrinsic scatter} & \multicolumn{4}{c}{Symmetric scaling}  \\ 
	\noalign{\smallskip}  
	Relation   & \multicolumn{2}{c}{$\alpha_{Y|X}$}	&	\multicolumn{2}{c}{$\beta_{Y|X}$}&	\multicolumn{2}{c}{$\gamma_z$}&	\multicolumn{2}{c}{$\sigma_{Y|X,0}$}&	\multicolumn{2}{c}{$\gamma_{\sigma_{Y|X}}$} & \multicolumn{2}{c}{$\alpha_{Y\textrm{-}X}$}	& \multicolumn{2}{c}{$\beta_{Y\textrm{-}X}$}\\ 
	\noalign{\smallskip}  
	\hline
	$\sigma_\mathrm{v}$-$M_{\mathrm{WL},200}$&	2.73&0.05	&   0.22&0.05&    -0.07&0.23&    0.07&0.01&    0.24&0.84& 2.71&0.06&  0.25&0.06	  \\
	$\lambda$-$M_{\mathrm{WL},200}$	&		1.50&0.06&   0.45&0.05&   0.41&0.51&   0.24&0.04&   -0.59&0.88&	1.33&0.10&	0.65&0.12  \\
	$L_\mathrm{X}$-$M_{\mathrm{WL},500}$			&	0.13&0.15&    1.29&0.14&    -2.00&0.63&   0.20&0.04&    0.23&0.80& -0.16&0.21&	1.65&0.24\\
	$Y_\mathrm{SZ}$-$M_{\mathrm{WL},500}$	&	-0.25&0.03&    1.00& 0.11&    -0.05& 0.54&   0.16&0.03&    -0.31&0.95&	-0.30&0.04& 	1.41&0.21   \\
	\hline
	\end{tabular}
\end{table*}

We analysed the scaling between mass and optical, X-ray, or SZ observables using the general regression scheme detailed in Section~\ref{sec_regr}. In fact, the uncertainties on the redshifts were assumed to be negligible and the factors $E_z$ were assumed to be known without errors. The Bayesian hierarchical model was implemented through JAGS.\footnote{JAGS (Just Another Gibbs Sampler) is a program for analysis of Bayesian hierarchical models using Markov Chain Monte Carlo simulation. It is publicly available at \url{http://mcmc-jags.sourceforge.net/}. An example of JAGS script used for the analysis can be found at \url{http://pico.bo.astro.it/\textasciitilde sereno/CoMaLit/JAGS/.}}

According to the notation of Section~\ref{sec_regr}, the $X$ variable is the logarithm (to base 10) of the observed weak lensing mass (computed at an over-density of either 200 in case of scaling of optical observables or 500 for observables related to the gas); the $Z$ variable is the logarithm of the unknown true mass; the observable $Y$ is the logarithm of either the galaxy velocity dispersion, the optical richness, the X-ray luminosity, or the spherical SZ signal multiplied by $E_z^{-\gamma_\mathrm{ss}}$. Any deviation of the parameter $\gamma_z$ from the null value implies a deviation from the self-similar evolution with redshift.

In the case of optical richness and SZ signal, we had to consider the correction for the Malmquist bias. The threshold value of the optical richness above which candidate clusters are included in the redMaPPer catalog was given by 20 times the scale factor at the cluster redshift \citep{ryk+al14}. According to the notation of Section~\ref{sec_cata_redmapper}, the threshold for the $i$-th cluster is
\beq
y_{\mathrm{th},i}=\log(20 S_{\mathrm{RM},i}).
\eeq

The limiting SZ flux of the {\it Planck} clusters was obtained by multiplying the minimum $\mathrm{SNR}(=4.5)$ by the uncertainty on $Y_{500}$ \citepalias{ser+al14_comalit_II}. In this case,
\beq
y_{\mathrm{th},i}=\log(4.5 E_{z,i}^{-2/3}D_\mathrm{A}^2(z_i) \delta_{Y_{500},i}).
\eeq

The results of the regression are summarised in Table~\ref{tab_scaling} for the scaling relations and in Table~\ref{tab_mass_func} for the mass functions. Table ~\ref{tab_scaling_MWL} summarises the results of the scaling of the observables versus the WL mass. Parameter degeneracies are illustrated as bi-dimensional contour plots in Figs.~\ref{fig_sigma_M_PDF},~\ref{fig_richness_M_PDF},~\ref{fig_LX_M_PDF}, and~\ref{fig_YSZ_M_PDF}. Figures~\ref{fig_sigma},~\ref{fig_richness},~\ref{fig_LX}, and ~\ref{fig_YSZ} show the scaling relation and the evolution of the completeness function for $\sigma_\mathrm{v}$-$M_{200}$, $\lambda$-$M_{200}$, $L_\mathrm{X}$-$M_{500}$, and $Y_\mathrm{SZ}$-$M_{500}$, respectively. 

To ease the comparison with theoretical predictions, we also computed the parameters of the the symmetric scaling relation (see Table~\ref{tab_scaling}). We did not require that $\beta_{Y\textrm{-}Z}$ is redshift independent. However, the slopes turned out to be constant within the errors. Slopes and intercepts of the symmetric relations in Tables~\ref{tab_scaling} and ~\ref{tab_scaling_MWL} were computed at the median redshifts of the samples.

We obtained significant constraints on the evolution with redshift of the scaling relations and of the mass functions. On the other hand, the uncertainties on the evolution of the intrinsic scatters are too large to come to any conclusion.

\subsection{Parameter degeneracy}

Most of the regression parameters are uncorrelated (see Figs.~\ref{fig_sigma_M_PDF}--\ref{fig_YSZ_M_PDF}). Since a significant percentage of the massive clusters is at high redshift, the time evolution can partially mimic the effects of the mass evolution (see the $\beta_{Y|Z}$-$\gamma_z$ panel). This degeneracy is most pronounced for the {\it Planck} selected clusters, see Fig.~\ref{fig_YSZ_M_PDF}, whose mass completeness limits steadily increases with redshift, see Fig.~\ref{fig_YSZ}. To obtain unbiased estimates of the scaling relations is then crucial to properly account for time-evolution and selection effects.

The estimate of the normalisation of the scaling relation, $\alpha_{Y|Z}$, is correlated with the slopes, $\beta_{Y|Z}$ and $\gamma_z$. Slope and normalisation are correlated to the intrinsic scatter too. The Pearson correlation coefficient between $\alpha_{Y|Z}$ ($\beta_{Y|Z}$) and $\sigma_{{Y|Z},0}$ is 0.38 (-0.38) for the $\sigma_\mathrm{v}$-$M_\mathrm{WL}$ sample, 0.72 (-0.75) for the $\lambda$-$M_\mathrm{WL}$ sample, 0.50 (-0.53) for the $L_\mathrm{X}$-$M_\mathrm{WL}$ sample, and -0.08 (-0.31) for the $Y_\mathrm{SZ}$-$M_\mathrm{WL}$ sample. 

The parameters characterising the scaling, i.e.,  $\alpha_{Y|Z}$, $\beta_{Y|Z}$, and $\gamma_z$, are not degenerate with the mass functions. Furthermore, as already noted in \citetalias{se+et14_comalit_I}, since $\sigma_{X|Z}$ and $\sigma_{Y|Z}$ spread the observed relation in orthogonal directions, they are nearly uncorrelated too. 

The only remaining significant degeneracy is among the parameters characterising the normalisation of the mean value of the mass function, $\bar{\mu}_Z$, and its evolution, parameterised in terms of $\gamma_{\mu_Z,D}$ and $\gamma_{\mu_Z}$. The evolution parameters of the (mean of the) mass function, $\gamma_{\mu_Z,D}$ and $\gamma_{\mu_Z}$, are degenerate too, see Eqs.~(\ref{eq_evo_16}) and (\ref{eq_bug_7}). In fact, the redshift dependence in small intervals can be modelled either in terms of $E_z$ or in terms of the distance.

\subsection{Scaling with the weak-lensing mass}

Results for the scalings with the WL mass are reported in Table ~\ref{tab_scaling_MWL}. In this case, the general regression scheme simplifies as described in Sec.~\ref{sec_regr}. Due to the intrinsic scatter of the WL mass with respect to the true mass, the conditional relation $O$-$M_{\mathrm{WL},\Delta}$ is systematically flatter than the corresponding $O$-$M_\Delta$, whereas the intrinsic scatter of the relation is larger \citepalias{se+et14_comalit_I,ser+al14_comalit_II}. On the other hand, results for the symmetric scaling relations are consistent.

Parameters of the $O$-$M_{\mathrm{WL},\Delta}$ are recovered with better accuracy but they cannot be used straight on in the conditional $O$-$M_\Delta$ to make predictions based on cosmological functions. In absence of a full regression procedure modelling the intrinsic scatter between true mass and proxy mass, at least some corrections should be used \citepalias{se+et14_comalit_I}.

\subsection{Self-similarity}

We found that scalings with mass ($\beta_{Y\textrm{-}Z} \sim \beta_\mathrm{ss}$) and redshift ($\gamma_z \sim 0$) are fully compatible with the self-similar behaviour, see Tables~\ref{tab_scaling} and \ref{tab_scaling_MWL}. The lone exception is the mass--X-ray luminosity relation, whose dependence with mass is slightly steeper than theoretical predictions even if compatible within the statistical uncertainty ($\beta_{Y\textrm{-}Z}=1.78\pm0.25$ vs. $\beta_\mathrm{ss}=4/3$), and whose evolution with redshift is preferentially negative ($\gamma_z=-1.8\pm0.8$). 

We tested that these general features of the $L_\mathrm{X}$-$M_\Delta$ relation do not depend on the different ways in which the X-ray luminosity can be measured. Bolometric luminosities estimated in a region from which the core is excised are supposed to be slightly less sensitive to the baryonic physics. Considering the $L_\mathrm{X,ce}$ from \citet{mau+al12}, we found $\beta_{Y\textrm{-}Z}=1.47\pm0.18$, $\gamma_z=-1.39\pm0.57$, and $\sigma_{Y|Z}=0.09\pm0.05$. As expected, the $L_\mathrm{X,ce}$-$M_\Delta$ relation is more tight, with a smaller intrinsic dispersion, and it can be determined with a better statistical accuracy. Nevertheless, the evolution in redshift is still negative and fully compatible with the result obtained for luminosities accounting for the core regions. The dependence with mass is slightly less pronounced but still steeper than the self-similar expectation.

We also evaluated the mass--X-ray luminosity relation in the soft band by considering the MCXC catalog, which comprises 193 clusters with measured WL mass. For the $L_\mathrm{X_{soft}}$-$M_\Delta$ relation we found $\beta_{Y\textrm{-}Z}=1.43\pm0.15$, which is steeper than the self-similar prediction $\beta_\mathrm{ss}=1$, and a negative redshift evolution which deviates from the expected $\gamma_\mathrm{ss}=2$ (see e.g. \citet{ett15} for a derivation of the self-similar predictions in the [0.1--2.4] keV band) by $\gamma_z=-1.80\pm0.59$. As expected for the MCXC catalog, whose luminosities were standardised from heterogeneous data sets, the intrinsic scatter is over-estimated, $\sigma_{Y|Z}=0.18\pm0.07$.

\subsection{Intrinsic scatters}

The intrinsic scatter of the weak lensing mass with respect to the true mass is estimated to be of the order of $\sim 25\pm10$ per cent. This is slightly larger but still compatible within errors with the scatter measured in \citet{man+al14} or in \citetalias{se+et14_comalit_I}, and with predictions based on numerical simulations \citep{ras+al12}. In \citetalias{ser+al14_comalit_II}, we noted that the scatter measured in heterogeneous samples, such as LC$^2$-{\it single}, may be overestimated due to not coherent formulated statistical uncertainties on weak lensing masses. In the case of the richness calibration, both estimated weak lensing scatter and related errors doubles, so that the difference in the central estimates of the scatter is not statistically significant.

We confirm that mass proxies based on velocity dispersions are among the most accurate (scatter of $\sim$14 per cent). We verified that mass proxies based on X-ray luminosity are noisier (scatter of $\sim$30 per cent) than those based on the integrated SZ effect ($\sim$25 per cent), in agreement with numerical simulations \citep{sta+al10}. The richness, with an intrinsic scatter of $\sim$40 per cent, may compete with the X-ray luminosity. The scatter in proxies based on the X-ray luminosity strongly depend on the measurement/analysis process. Luminosities estimated in core-excised regions are less scattered ($\sim$20 per cent), whereas inconsistencies in the measurement process and not uniform data-sets can boost the scatter up to $\sim 40$ per cent.

The catalog of velocity dispersion used to study the $\sigma_\mathrm{v}$-$M_\mathrm{200}$ relation is heterogeneous which might bias the relation and inflate the estimated intrinsic scatter. Firstly, the central estimates of the velocity dispersion were estimated in the different source papers with different methods. However, these statistical differences are very small. \citet{rue+al14} presented two independent measurements of $\sigma_\mathrm{v}$, based either on the measured gapper scale or the biweight dispersion. The two estimates are very well consistent, with a distribution of relative differences whose centre deviates from zero by less than one per cent and whose scatter is less than 2 per cent. 

Secondly, systematic differences in the measurements by independent groups might play a role. We checked that results based on the merged catalogs are fully consistent with those from single source catalogs with homogeneous estimates of  $\sigma_\mathrm{v}$. 31 clusters from \citet{gi+me01} have measured WL mass. For these clusters, we found $\alpha_{Y|Z}=2.74\pm0.33$, $\beta_{Y|Z}=0.22\pm0.34$, $\gamma_z=0.04\pm0.54$, and $\sigma_{Y|Z,0}=0.07\pm0.03$. For the 26 clusters from \citet{rin+al13} with measured WL mass, we found $\alpha_{Y|Z}=2.72\pm0.27$, $\beta_{Y|Z}=0.26\pm0.29$, $\gamma_z=-0.56\pm1.23$, and $\sigma_{Y|Z,0}=0.07\pm0.02$.

\subsection{Completeness function}

\begin{figure}
\begin{tabular}{c}
\resizebox{\hsize}{!}{\includegraphics{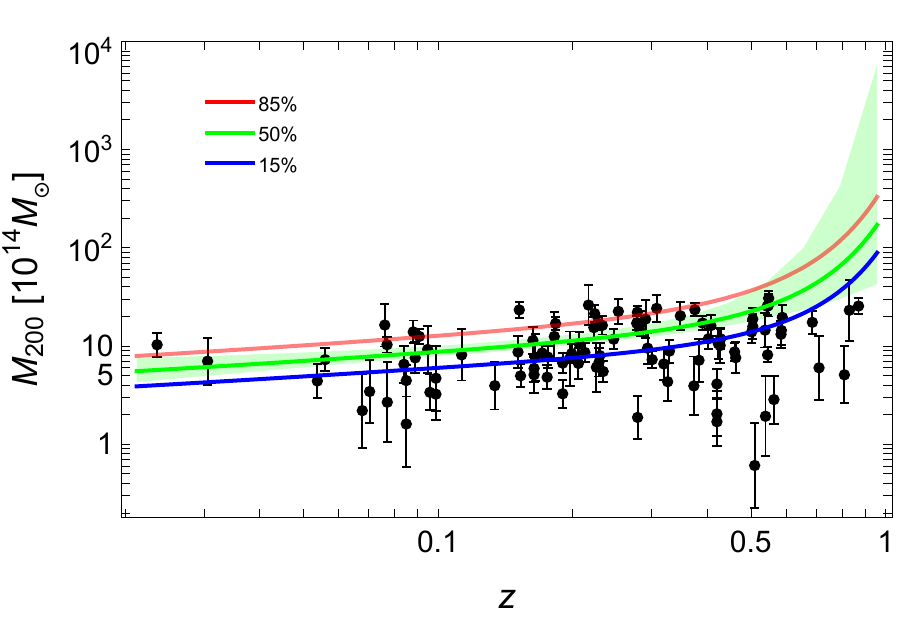}} \\
\noalign{\smallskip}  
\resizebox{\hsize}{!}{\includegraphics{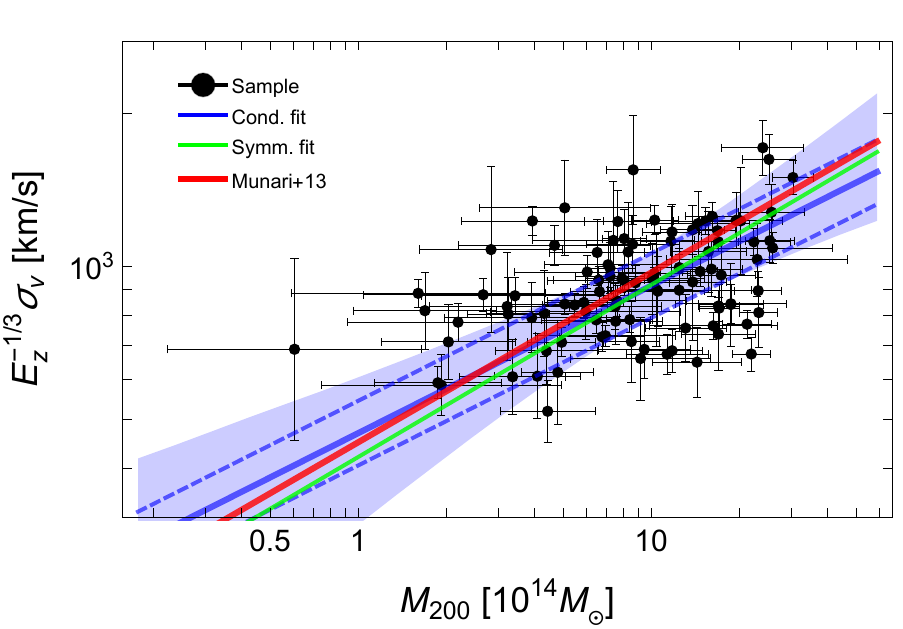}} \\
\end{tabular}
\caption{{\it Top panel}: completeness function of the clusters from the WL-velocity dispersion sample. Weak lensing masses $M_{\mathrm{WL},200}$ are shown as a function of redshift (black points). The full lines plot the value of the true mass $M_{200}$ where a given completeness level is reached as a function of the redshift. From top to bottom, the red, green, and blue lines show the 85, 50, and 15 per cent completeness levels, respectively. The shaded green region encloses the 68 per cent confidence region around the 50 per cent completeness level due to uncertainties on the mass function parameters. Note that the completeness is a function of the true mass, whereas the points refer to the weak lensing masses, which are scattered with respect to the true mass. {\it Bottom panel:} scaling between velocity dispersion and mass, $M_{200}$. The black points mark the data (weak lensing mass and redshift evolved velocity dispersion), the blue (green) line represent the conditional (symmetric) scaling relation fitted to the data (true mass versus redshift evolved velocity dispersion). The dashed blue lines show the median scaling relation (full blue line) plus or minus the intrinsic scatter. The shaded blue region encloses the $68$ per cent confidence region around the median relation due to uncertainties on the scaling parameters. The red line represents the theoretical prediction based on \citet{mun+al13} at the median redshift of the sample. Masses are in units of $10^{14}M_\odot$.}
\label{fig_sigma}
\end{figure}

\begin{figure}
\begin{tabular}{c}
\resizebox{\hsize}{!}{\includegraphics{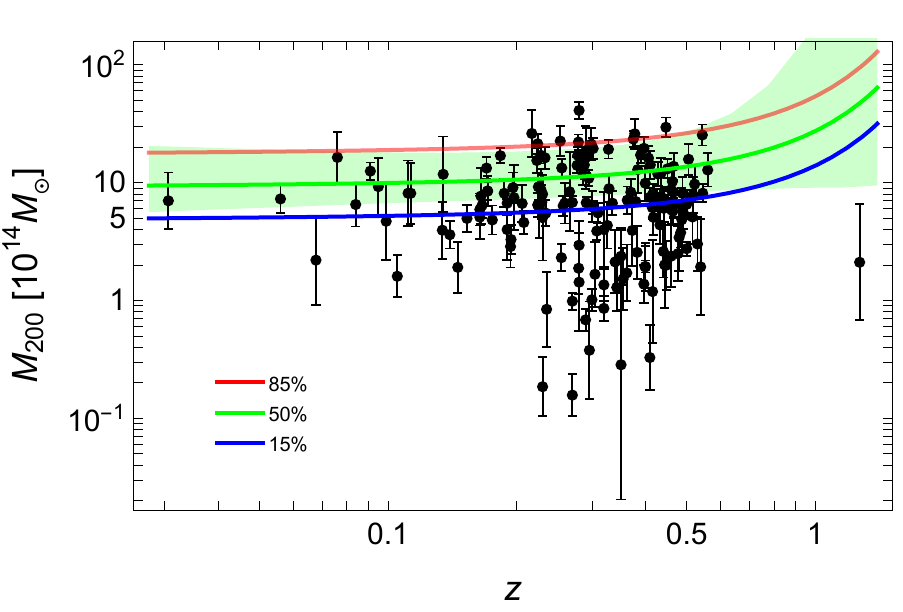}} \\
\noalign{\smallskip}  
\resizebox{\hsize}{!}{\includegraphics{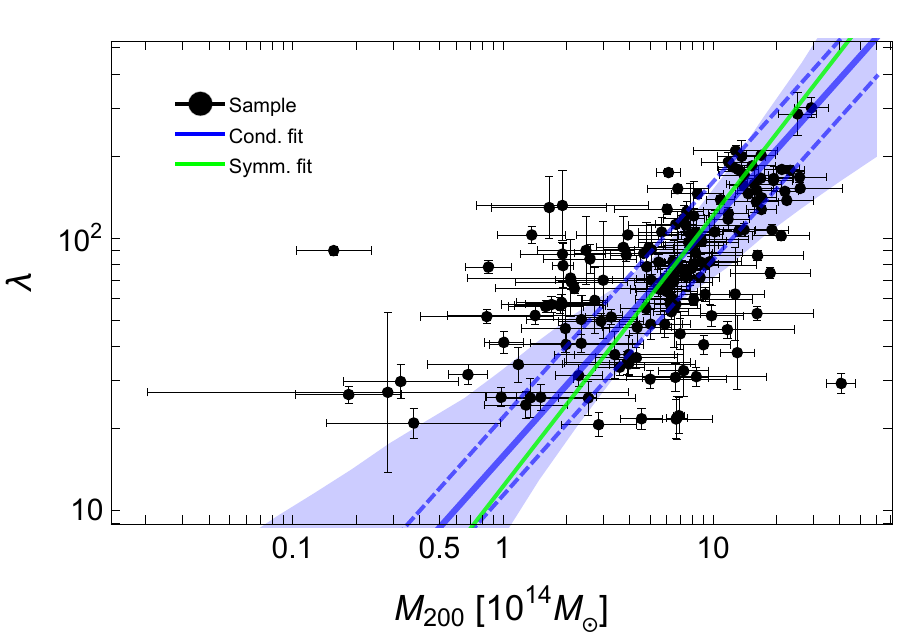}} \\
\end{tabular}
\caption{{\it Top panel}: completeness function of the clusters from the WL-optical richness sample. Lines and points are as in the top panel of Fig.~\ref{fig_sigma}. {\it Bottom panel}: scaling between optical richness and mass, $M_{200}$. Black, blue, and green graphics are as in the bottom panel of Fig.~\ref{fig_sigma}.}
\label{fig_richness}
\end{figure}

\begin{figure}
\begin{tabular}{c}
\resizebox{\hsize}{!}{\includegraphics{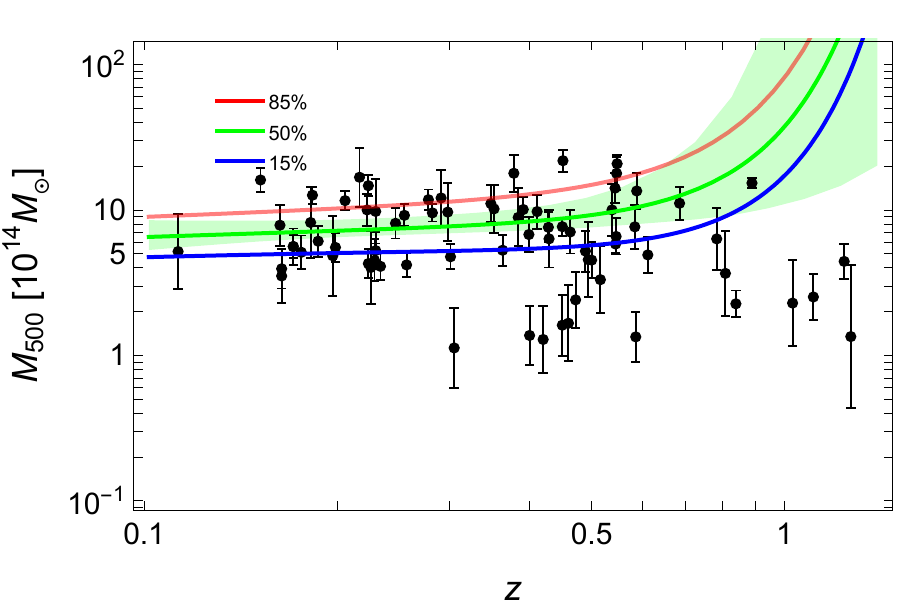}} \\
\noalign{\smallskip}  
\resizebox{\hsize}{!}{\includegraphics{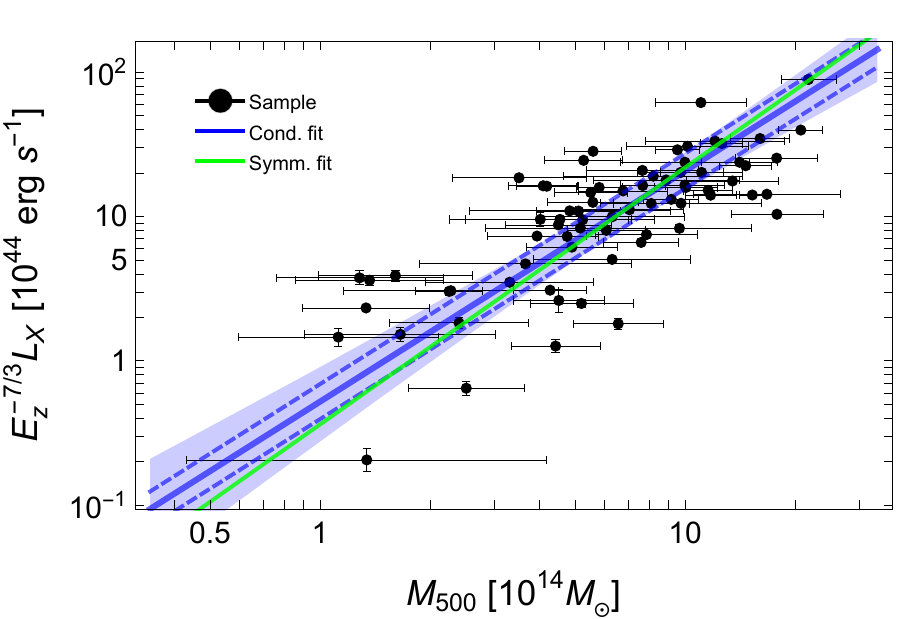}} \\
\end{tabular}
\caption{{\it Top panel}: completeness function of the clusters from the WL-X-ray luminosity sample. Lines and points are as in the top panel of Fig.~\ref{fig_sigma}. {\it Bottom panel}: scaling between (self-similarly redshift evolved) X-ray bolometric luminosity and mass $M_{500}$. Black, blue, and green graphics are as in the bottom panel of Fig.~\ref{fig_sigma}. 
}
\label{fig_LX}
\end{figure}

\begin{figure}
\begin{tabular}{c}
\resizebox{\hsize}{!}{\includegraphics{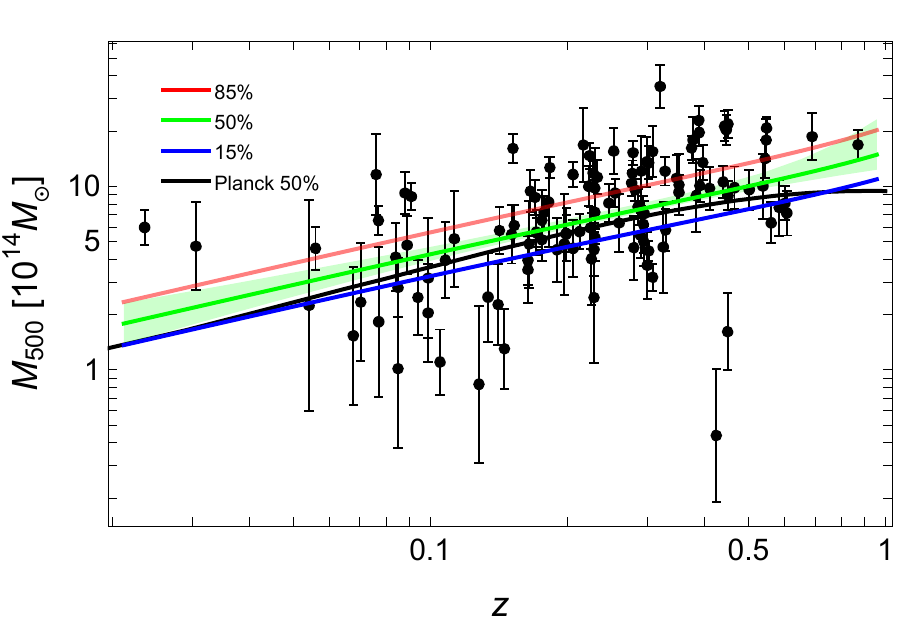}} \\
\noalign{\smallskip}  
\resizebox{\hsize}{!}{\includegraphics{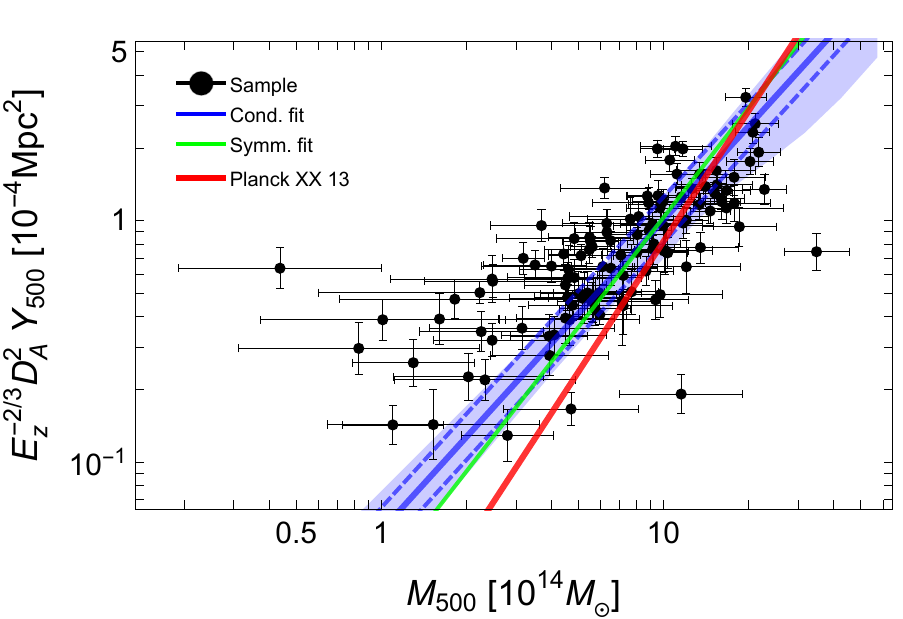}} \\
\end{tabular}
\caption{{\it Top panel}: completeness function of the clusters from the WL-SZ sample. Lines and points are as in the top panel of Fig.~\ref{fig_sigma}. The black line represents the 50 per cent completeness limit of the full SZ sample as derived in \citet{planck_2013_XXIX} from the average noise over the sky for the MMF3 algorithm. {\it Bottom panel}: scaling between (self-similarly redshift evolved) spherical SZ flux and mass, $M_{500}$. Black, blue, and green graphics are as in the bottom panel of Fig.~\ref{fig_sigma}. The red line represents the result from \citet{planck_2013_XX} at the median redshift of the sample.}
\label{fig_YSZ}
\end{figure}

The evolution of the completeness function is obtained through the analysis of the redshift dependence of the distribution of selected true masses. In Figs.~\ref{fig_sigma},~\ref{fig_richness},~\ref{fig_LX}, and ~\ref{fig_YSZ}, we plotted the 15, 50, and 85 per cent completeness limit as a function of the redshift. The completeness was approximated as a complementary error function, see App.~\ref{app_mass}, whose scale and dispersion were derived from the parameters of the fitted mass function through Eqs.~(\ref{eq_pmu_7}) and~(\ref{eq_pmu_8}). 

For the samples of clusters with weak lensing mass and velocity dispersion, SZ flux, or X-ray luminosity, the larger the redshift the larger the mass at a given completeness limit, see Figs.~\ref{fig_sigma},~\ref{fig_LX}, and ~\ref{fig_YSZ}. This is not the case for the WL clusters in the redMaPPer catalog, when the limits are nearly redshift independent, see Fig.~\ref{fig_richness}. The apparent spike in some completeness functions at high redshift, see Figs.~\ref{fig_sigma},~\ref{fig_richness}, and~\ref{fig_LX}, is not statistically significant and more regular evolutions are fully compatible with our results.

Few high redshift clusters lie above the 85 per cent completeness limit. This is expected since high mass clusters are very rare at high redshifts and few of them have measured WL mass.

Even though the redMaPPer catalog of optical richnesses \citep{ryk+al14} and the {\it Planck} catalogs \citep{planck_2013_XXIX} are statistically complete, we do not expect to exactly recover the selection function of the full catalogs. In fact, the sub-samples of clusters with measured weak lensing masses may be biased with respect to the full catalogs.

The derived completeness functions refer to the subsample of the cluster in the catalogs with measured WL mass. The catalog of WL masses is heterogeneous which makes the studied subsamples heterogeneous too, even in the case of parent catalogs constructed with well defined selection functions. However, despite the fact that we do not expect that the completeness function of the subsamples with WL masses strictly resembles the completeness of the parent catalog, some similarities are still in place. The redshift evolution of the derived completeness limits of the SZ flux-WL catalog follows that of the {\it Planck} clusters, see the upper panel of Fig.~\ref{fig_YSZ}. The shift in normalisation reflects the fact that masses used in \citet{planck_2013_XXIX} to derive the average limit were based on the $Y_z$ proxy which severely under-estimates the true masses (\citealt{lin+al14};\citetalias{ser+al14_comalit_II}). Furthermore, the subsample with WL masses under-represents the clusters with low SNR just above the threshold \citepalias{ser+al14_comalit_II}. The larger mean SNR of the subsample determines a larger average limit, as we observed.

The flatness of the completeness limits of the optical richness-WL subsample (see upper panel of Fig.~\ref{fig_richness}) is also connected to the properties of the parent sample. The redMaPPer catalog is in fact nearly complete at $z\ls 0.3$ and $\lambda \gs 30$ \citep{ryk+al14}.

\subsection{Mass function}

\begin{figure}
\begin{tabular}{c}
\includegraphics[width=7.cm]{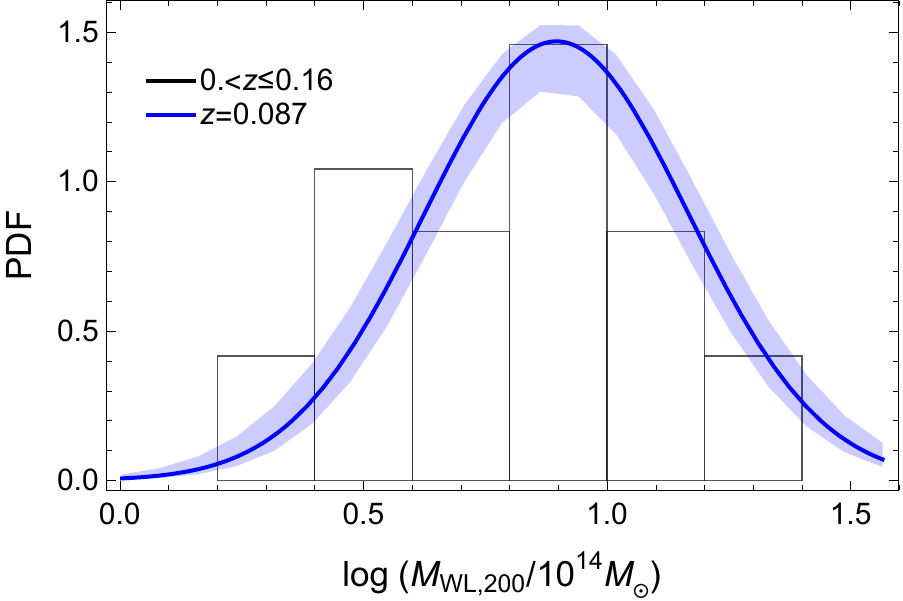} \\
\includegraphics[width=7.cm]{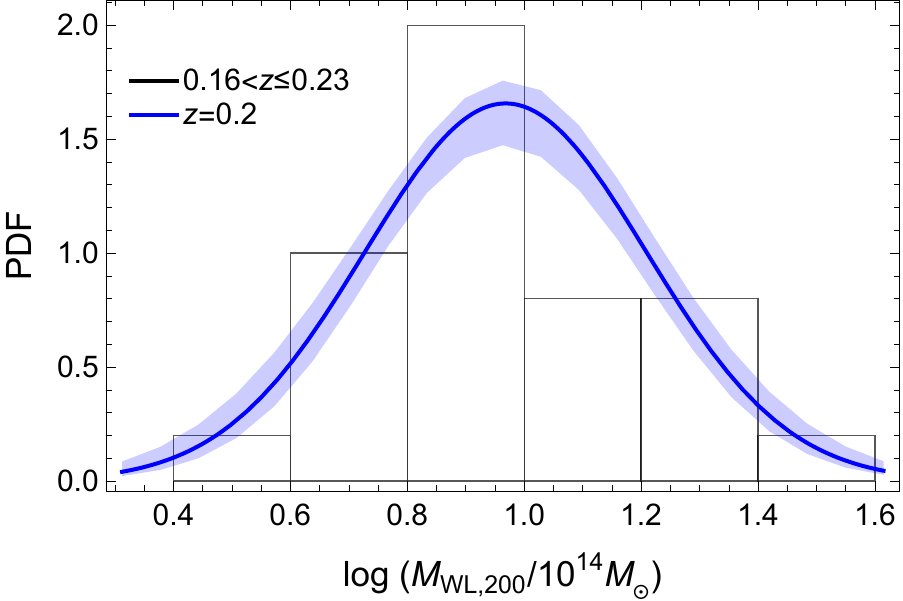} \\
\includegraphics[width=7.cm]{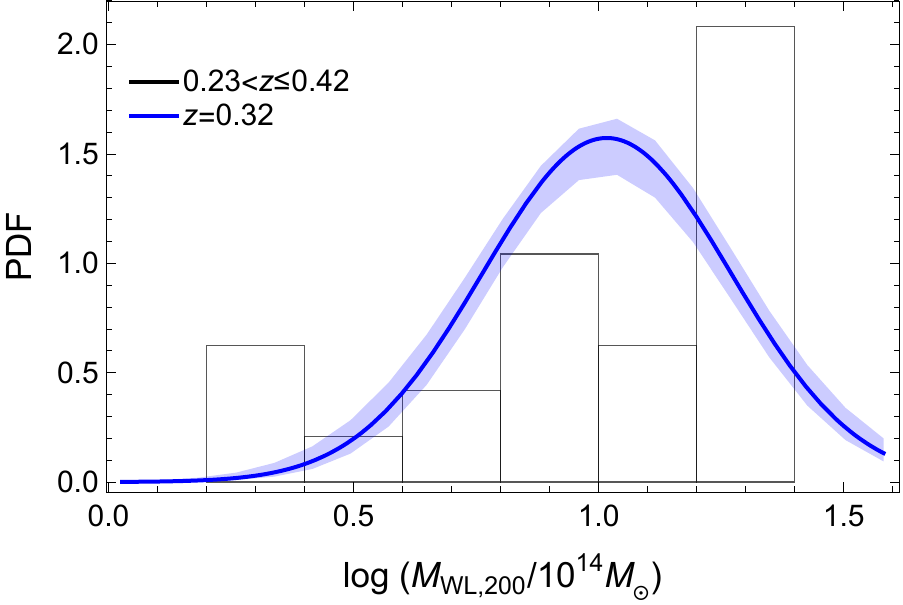} \\
\includegraphics[width=7.cm]{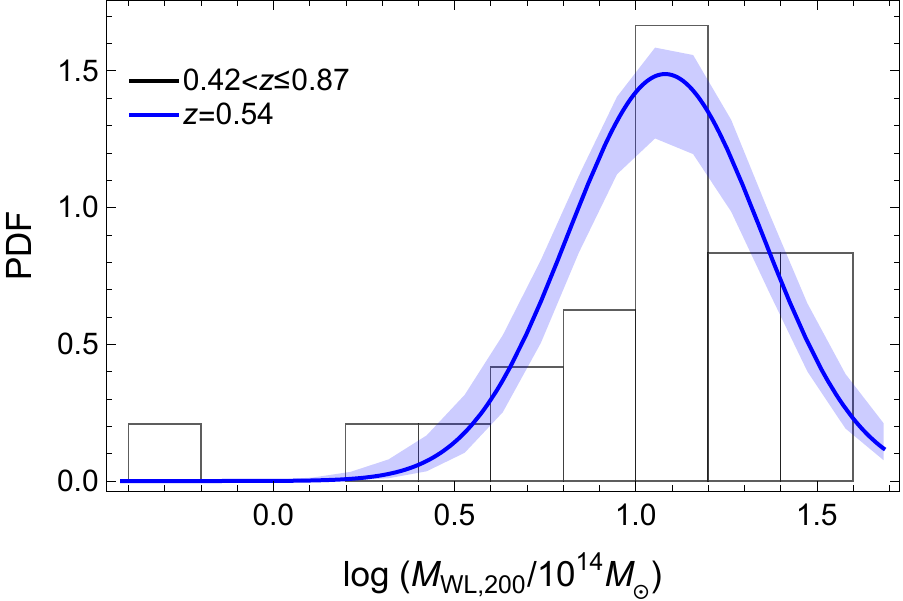} \\
\end{tabular}
\caption{Mass function of the clusters from the $\sigma_\mathrm{v}$-$M_\mathrm{WL}$ sample in four redshift bins. The black histogram groups the observed WL masses. The blue line is the normal approximation estimated from the regression at the median redshift. The shaded blue region encloses the 68 per cent confidence region around the median relation due to uncertainties on the parameters of the mass function. The mass function for the observed WL masses is estimated from the fit result, i.e., the mass function of the true masses, by smoothing the prediction with a Gaussian whose variance is given by the quadratic sum of the intrinsic scatter of the (logarithmic) WL mass with respect to the true mass and the median observational uncertainty on the WL mass. Redshift increases from the top to the bottom. The median redshift and the redshift bins are indicated in the legends of the respective panels.}
\label{fig_sigma_M200_histo}
\end{figure}

\begin{figure}
\begin{tabular}{c}
\includegraphics[width=7.cm]{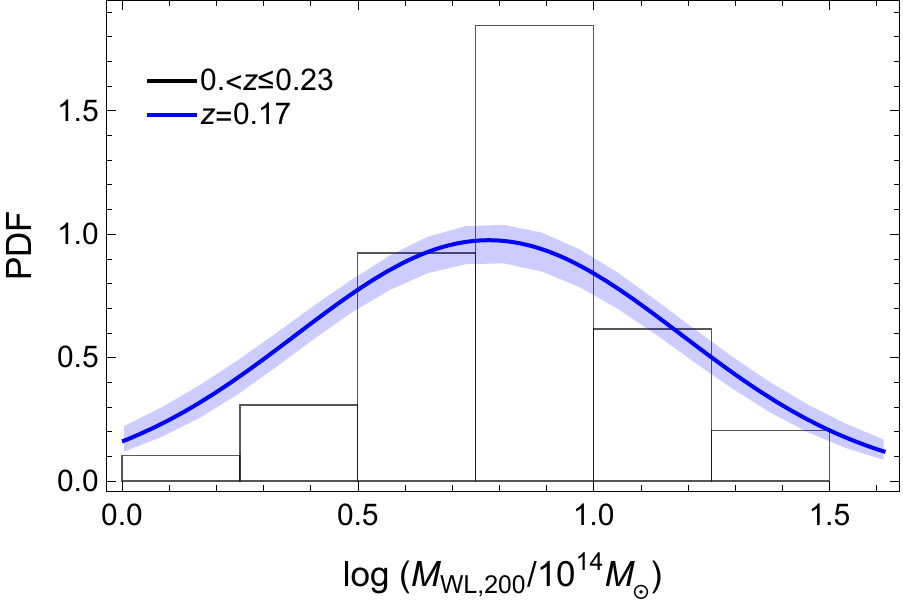} \\
\includegraphics[width=7.cm]{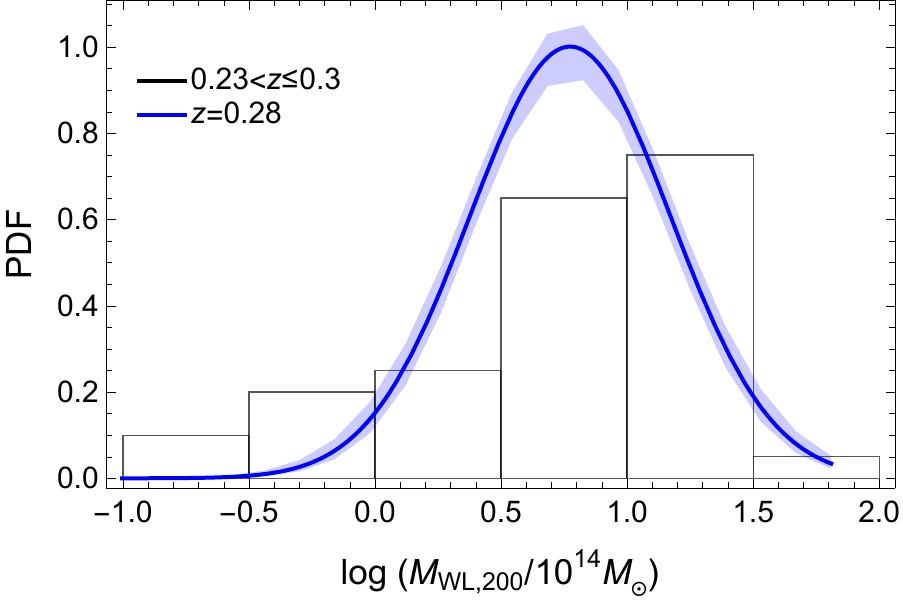} \\
\includegraphics[width=7.cm]{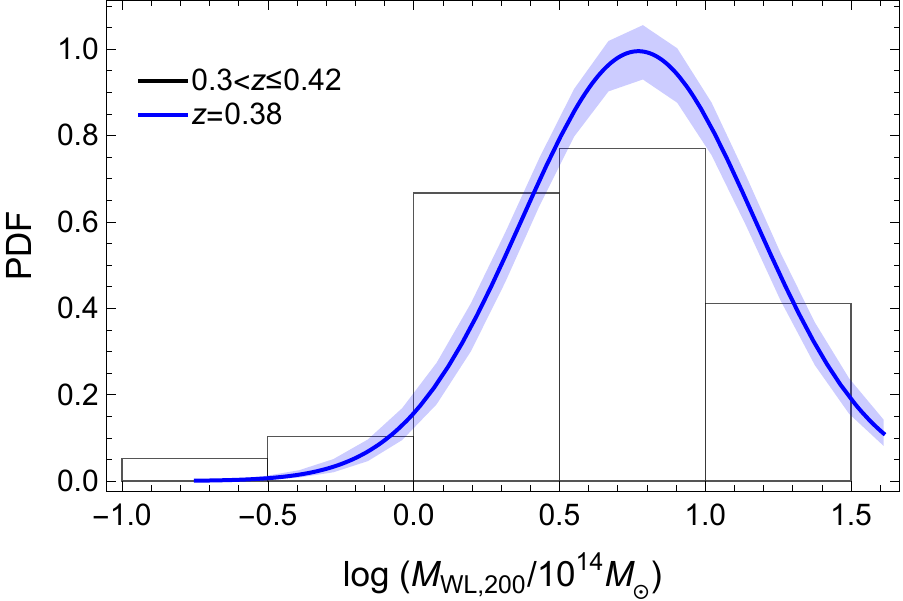} \\
\includegraphics[width=7.cm]{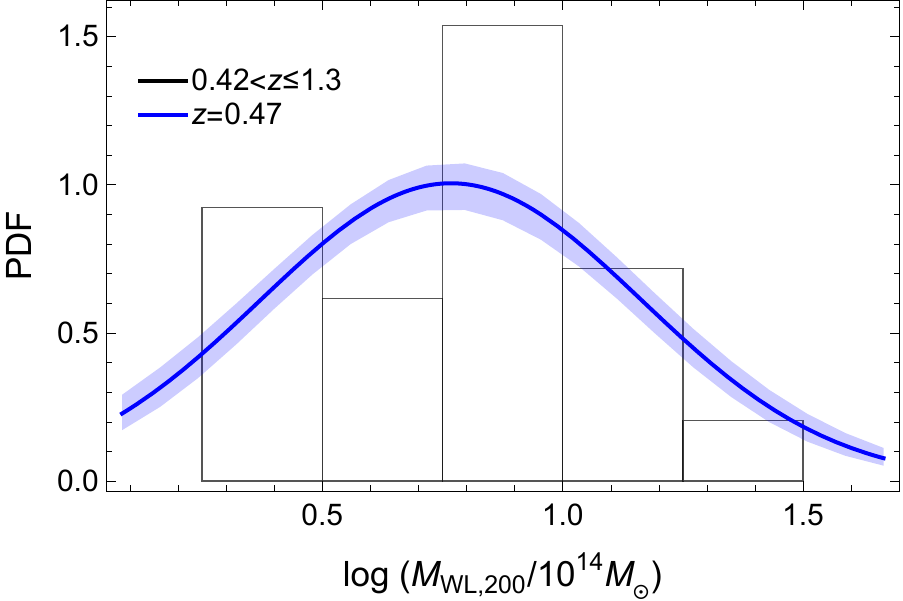} \\
\end{tabular}
\caption{Mass function of the clusters from the optical richness-WL sample in four redshift bins. Lines and conventions are as in Fig.~\ref{fig_sigma_M200_histo}.}
\label{fig_richness_M200_histo}
\end{figure}

\begin{figure}
\begin{tabular}{c}
\includegraphics[width=7.cm]{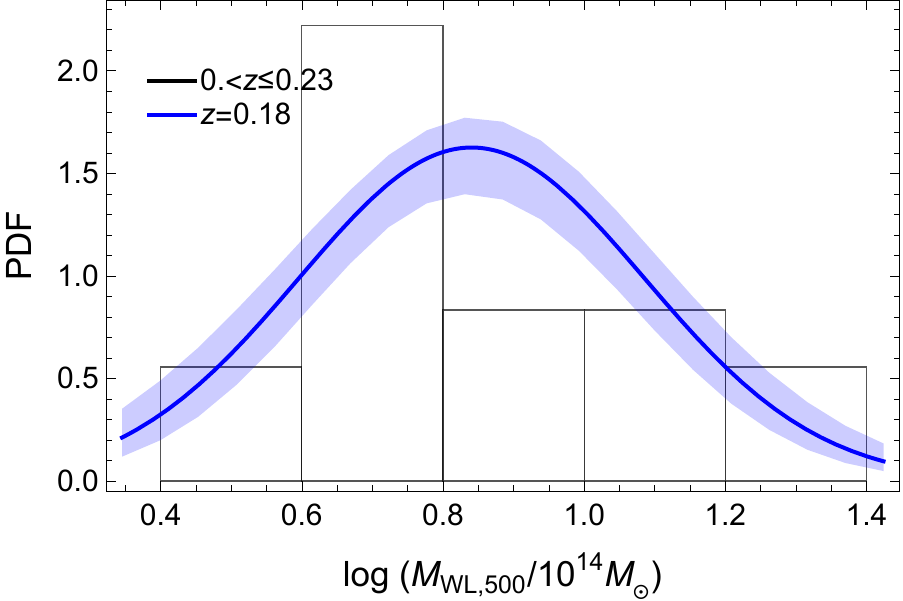} \\
\includegraphics[width=7.cm]{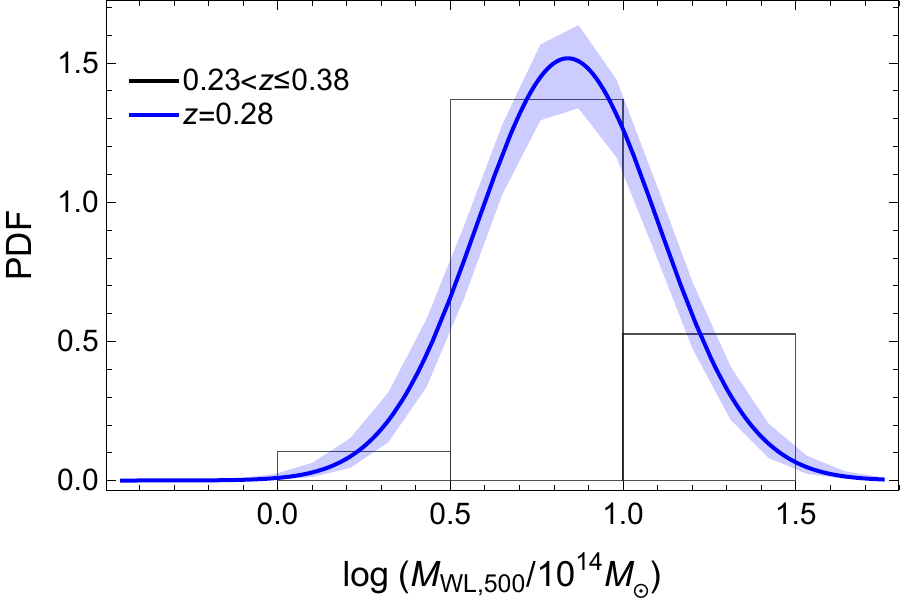} \\
\includegraphics[width=7.cm]{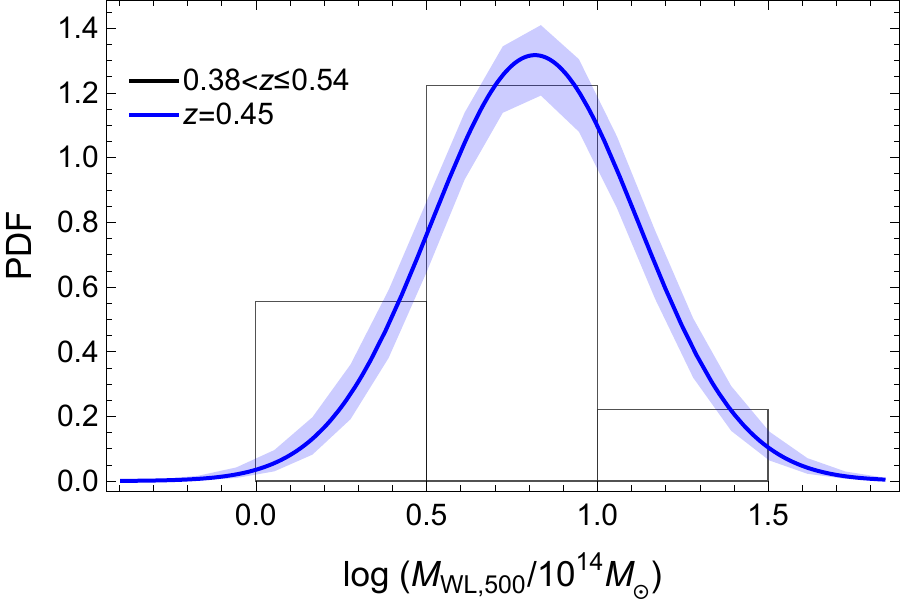} \\
\includegraphics[width=7.cm]{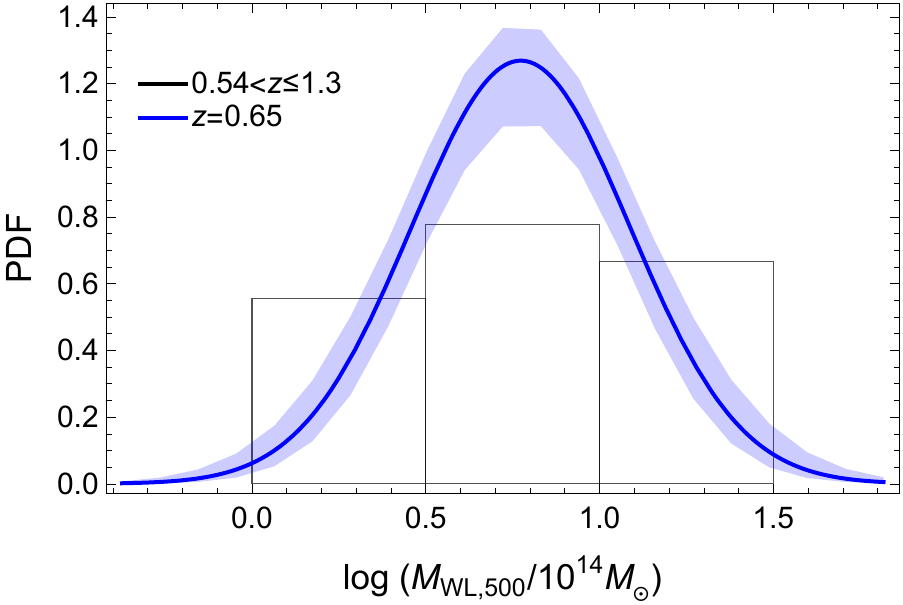} \\
\end{tabular}
\caption{Mass function of the clusters from the X-ray luminosity-WL sample in four redshift bins. Lines and conventions are as in Fig.~\ref{fig_sigma_M200_histo}.}
\label{fig_LX_M500_histo}
\end{figure}

\begin{figure}
\begin{tabular}{c}
\includegraphics[width=7.cm]{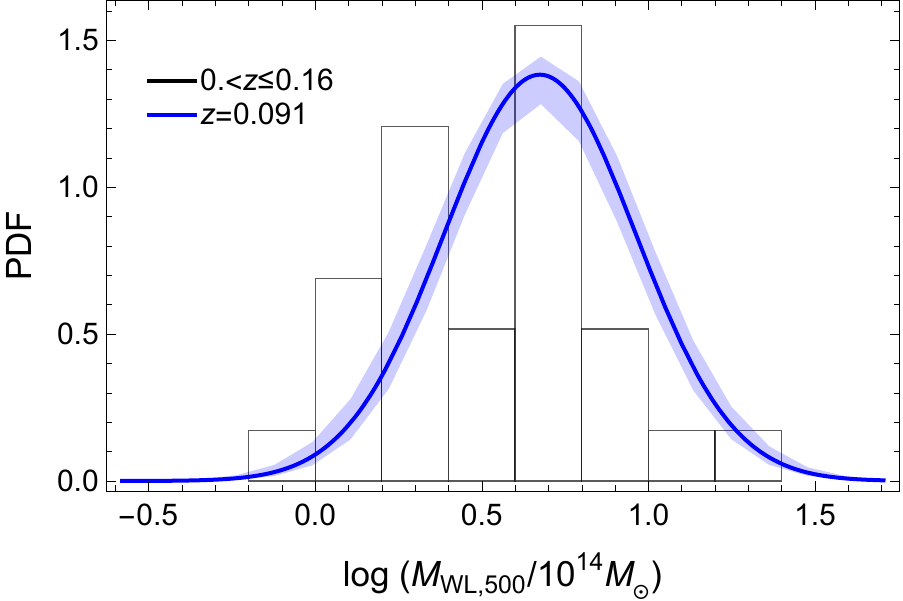} \\
\includegraphics[width=7.cm]{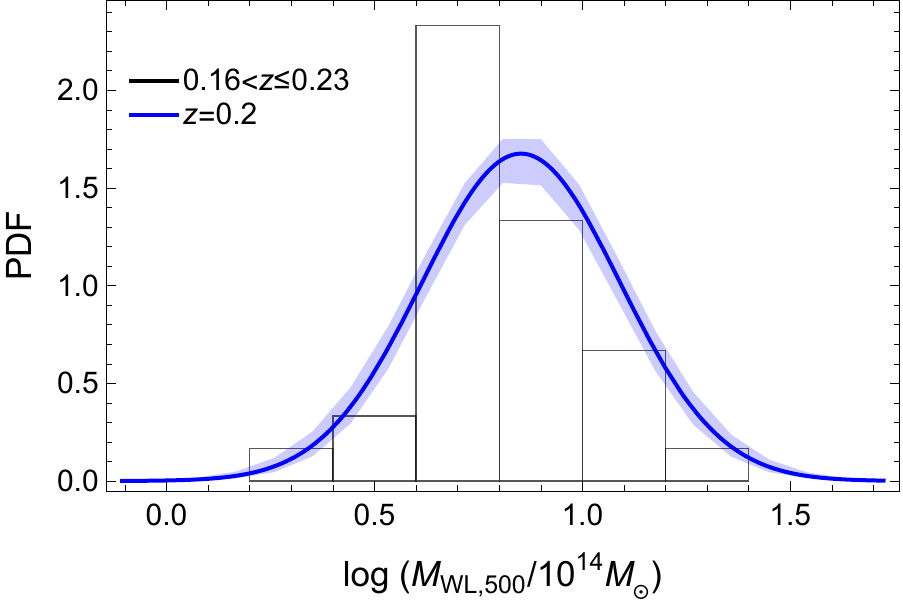} \\
\includegraphics[width=7.cm]{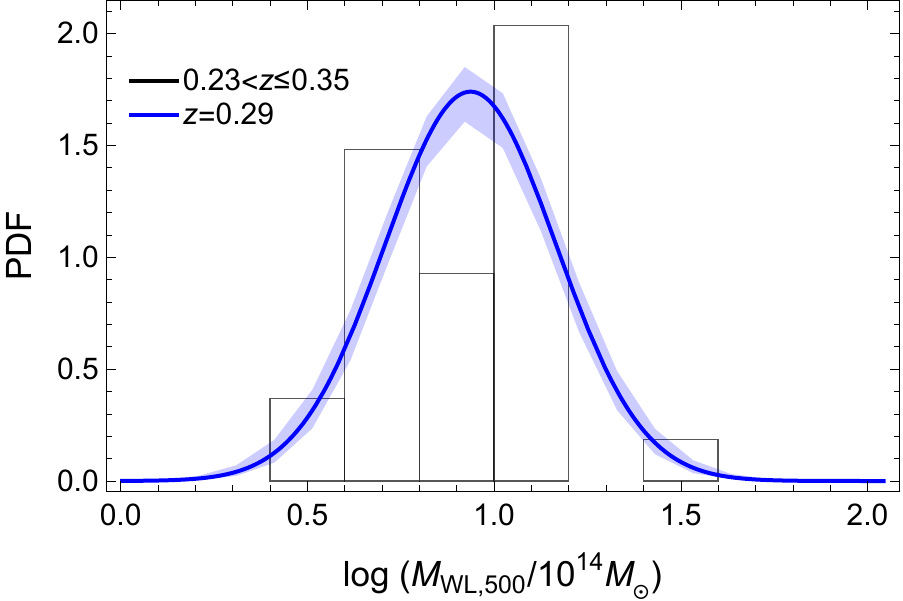} \\
\includegraphics[width=7.cm]{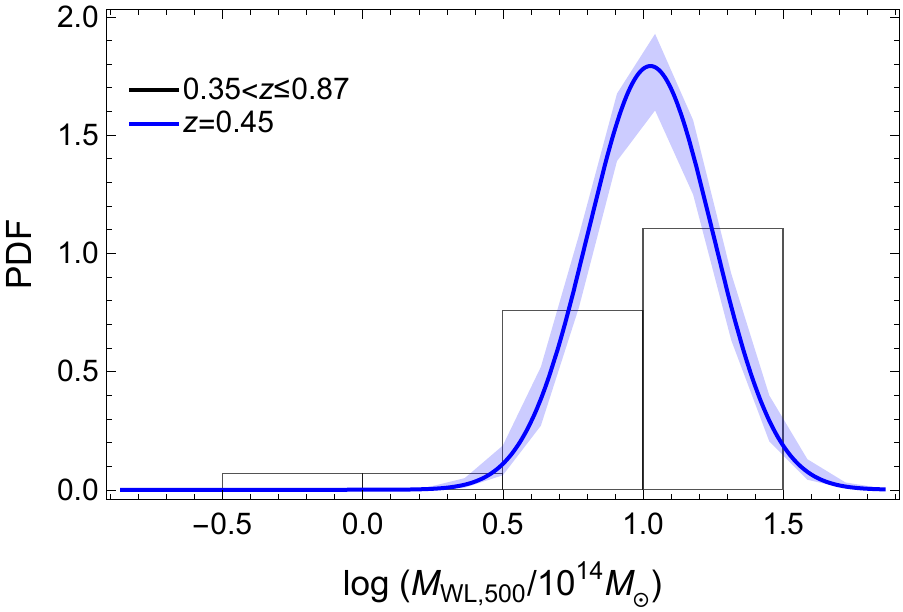} \\
\end{tabular}
\caption{Mass function of the clusters from the SZ-WL sample in four redshift bins. Lines and conventions are as in Fig.~\ref{fig_sigma_M200_histo}.}
\label{fig_YSZ_M500_histo}
\end{figure}

The modelling with a redshift-evolving Gaussian function, see Eqs.~(\ref{eq_bug_6},~\ref{eq_bug_7}) and (\ref{eq_bug_8}), is functional as far as the mass distribution is fairly unimodal and the redshift evolution is smooth. In the case of a cluster sample selected through a hard cut in the observable, these assumptions are well justified, see Sec.~\ref{sec_evol} and App.~\ref{app_mass}. However, the scheme is flexible enough to work even with heterogeneous sample since detections and measurements are generally limited by observational thresholds.

As far as the distribution is unimodal, a simple parameterisation of the distribution of the covariate variable in terms of a single Gaussian function can determine unbiased estimates of the scaling relation. Results are full consistent with more complex parameterisations adopting mixtures of Gaussians (\citealt{kel07};\citetalias{ser+al14_comalit_II}).

The observed WL mass functions in different redshift bins are well reproduced by the regression model. Distributions for the velocity dispersion--, optical richness--, X-ray luminosity--, and SZ flux--WL samples are shown in Figs.~\ref{fig_sigma_M200_histo}, \ref{fig_richness_M200_histo}, \ref{fig_LX_M500_histo}, and \ref{fig_YSZ_M500_histo}, respectively.  The regression model computes the distribution of the true masses. To compare them with the observed distribution of WL masses, we had to smooth the distribution firstly with a Gaussian function whose standard deviation is the intrinsic scatter between true and proxy mass, which is computed by the regression too, and then with a Gaussian whose dispersion is given by the observational uncertainty on the WL masses. We considered the median uncertainty in the redshift bin.

\subsection{Predictions for the CLASH sample}

\begin{table*}
\caption{Predicted line-of-sight velocity dispersions of the CLASH clusters. WL masses (col.~3) and redshifts (col.~2) are from the LC$^2$-{\it all}. The quoted uncertainty on $\sigma_\mathrm{v}$ is statistical (col.~4), including the contribution from the uncertainty in the mass estimates ($\delta\sigma_\mathrm{v}^\mathrm{mass}$, col.~5), from the uncertainties in the scaling relation parameters ($\delta\sigma_\mathrm{v}^\mathrm{SR}$, col.~6), and from the intrinsic scatter in the $\sigma_\mathrm{v}$-$M_\mathrm{WL}$ scaling relation and the related uncertainty ($\delta\sigma_\mathrm{v}^\mathrm{scat}$, col.~7).}
\label{tab_clash}
\centering
\begin{tabular}[c]{l  c r@{$\,\pm\,$}l r@{$\,\pm\,$}l r r r}
	\hline
	Name& $z$&		\multicolumn{2}{c}{$M_{\mathrm{WL},200}$}	&	\multicolumn{2}{c}{$\sigma_\mathrm{v}$} & $\delta\sigma_\mathrm{v}^\mathrm{mass}$&  $\delta\sigma_\mathrm{v}^\mathrm{SR}$ &  $\delta\sigma_\mathrm{v}^\mathrm{scatter}$\\
	& &		\multicolumn{2}{c}{$[10^{14}M_\odot]$}	&	\multicolumn{2}{c}{$[\mathrm{km/s}]$} & $[\mathrm{km/s}]$& $[\mathrm{km/s}]$& $[\mathrm{km/s}]$\\
	\hline
ABELL 383	&	0.187	&	8.1	&	2.2	&	958	&	164	&	63	&	25	&	153	\\
ABELL 209	&	0.206	&	17.6	&	3.0	&	1145	&	193	&	64	&	48	&	187	\\
ABELL 2261	&	0.224	&	21.2	&	4.1	&	1204	&	206	&	79	&	60	&	200	\\
RX J2129.3+0005	&	0.234	&	5.5	&	1.4	&	902	&	157	&	61	&	35	&	147	\\
ABELL 611	&	0.288	&	14.1	&	3.9	&	1139	&	197	&	77	&	34	&	183	\\
MS 2137.3-2353	&	0.313	&	12.4	&	4.8	&	1123	&	206	&	100	&	29	&	180	\\
RXC J2248.7-4431	&	0.348	&	20.2	&	6.7	&	1274	&	231	&	108	&	55	&	210	\\
MACS J1931.8-2635	&	0.352	&	14.7	&	6.4	&	1192	&	225	&	119	&	37	&	192	\\
MACS J1115.8+0129	&	0.352	&	15.5	&	3.4	&	1205	&	204	&	70	&	39	&	196	\\
RX J1532.9+3021	&	0.363	&	7.1	&	1.9	&	1022	&	177	&	70	&	35	&	167	\\
MACS J1720.3+3536	&	0.391	&	13.4	&	3.1	&	1195	&	204	&	70	&	36	&	192	\\
MACS J0416.1-2403	&	0.396	&	10.3	&	2.2	&	1130	&	191	&	62	&	32	&	182	\\
MACS J0429.6-0253	&	0.399	&	9.4	&	3.0	&	1109	&	197	&	85	&	33	&	180	\\
MACS J1206.2-0847	&	0.44	&	15.8	&	3.6	&	1272	&	219	&	78	&	46	&	208	\\
MACS J0329.6-0211	&	0.45	&	9.9	&	1.5	&	1156	&	193	&	55	&	39	&	189	\\
RX J1347.5-1145	&	0.451	&	29.3	&	6.1	&	1465	&	260	&	110	&	88	&	250	\\
MACS J1149.5+2223	&	0.544	&	25.2	&	5.2	&	1495	&	265	&	108	&	85	&	257	\\
MACS J0717.5+3745	&	0.548	&	30.5	&	4.9	&	1561	&	278	&	114	&	100	&	272	\\
MACS J0647.7+7015	&	0.584	&	13.1	&	4.2	&	1326	&	245	&	113	&	63	&	224	\\
MACS J0744.9+3927	&	0.686	&	17.3	&	4.7	&	1497	&	278	&	127	&	90	&	261	\\
	\hline
	\end{tabular}
\end{table*}

We can now make use of the constraints on the investigated scaling relations to obtain robust predictions on the observed properties of the galaxy clusters that are part of the CLASH program \citep[Cluster Lensing And Supernova survey with Hubble,][]{pos+al12}, which provides one of the samples best studied at different wavelengths.

The scaling relation $\sigma_\mathrm{v}$-$M_{200}$ can be used to predict the results for the ongoing measurements of the velocity dispersions acquired as part of the extension with optical spectroscopic data of the objects in the CLASH sample (e.g., the VLT-VIMOS Large Program of 230 hours to carry out a panoramic spectroscopic survey of the 14 southern CLASH clusters). The CLASH-VLT program aims at obtaining redshift measurements for 400-600 cluster members and 10-20 lensed multiple images in each cluster field \citep{biv+al13}. Additional observations from northern facilities, i.e., the MMT/Hectospec \citep{rin+al13}, will complement the program.

For these predictions, we performed a new regression excising from the sample the eight CLASH clusters previously included. Since we are interested in predicting the velocity dispersion given the weak lensing mass, we considered the conditional $\sigma_\mathrm{v}$-$M^\mathrm{WL}_{200}$ relation rather than the true mass--velocity dispersion $\sigma_\mathrm{v}$-$M_{200}$. The results of the regression were in full agreement with the regression of the full sample: $\alpha_{Y|X}=2.75\pm0.05$; $\beta_{Y|X}=0.22\pm0.05$; $\gamma_z=-0.02\pm0.24$; $\sigma_{Y|X,0}=0.07\pm0.01$; $\gamma_{\sigma_{Y|X}}=0.07\pm0.91$. Alternatively, we might have used the $\sigma_\mathrm{v}$-$M_{200}$ scaling considering the additional source of error given by the intrinsic scatter between the known weak lensing masses and the unknown true masses. As a general remark, if we have weak lensing masses we cannot make predictions by plugging them in scaling relations which compare for example the observable to the hydrostatic mass.

We considered for the CLASH clusters the weak lensing masses reported in LC$^2$-{\it all} \citepalias{ser14_comalit_III} and based on \citet{ume+al14}, who performed a combined analysis of shear and magnification. The velocity dispersions based on the $\sigma_\mathrm{v}$-$M_{\mathrm{WL},{200}}$ relation are listed in Table~\ref{tab_clash}. The main source of statistical uncertainty on the predictions is due to the intrinsic scatter of the relation, which abundantly tops the uncertainties due to the propagated error on the WL mass or due to the uncertainties in the scaling relation parameters.

The prediction for MACS J1206.2-0847 compares well with the first measurement from CLASH-VLT \citep[ $\sigma_\mathrm{v}=1087^{+53}_{-55}~\mathrm{km/s}$]{biv+al13}. The HeCS covered two clusters later on analysed in \citet{ume+al14}, i.e., A2261 and RXJ2129. The prediction for RXJ2129 is in excellent agreement with the measurements \citep[ $\sigma_\mathrm{v}=858^{+71}_{-57}~\mathrm{km/s}$]{rin+al13}. On the other hand, the prediction slightly exceeds the observed velocity dispersion of A2261 \citep[ $\sigma_\mathrm{v}=780^{+78}_{-60}~\mathrm{km/s}$]{rin+al13}, even though the small discrepancy is fully covered by the uncertainty due to the intrinsic dispersion of the scaling.

\section{Comparisons}
\label{sec_comparisons}

In this section, we compare our results to theoretical predictions or previous estimates.

\subsection{Theoretical predictions}
\label{sec_theo}

We first compare our results to predictions based either on theoretical models of structure formation or on numerical/hydro-dynamical simulations.

\subsubsection{$L_\mathrm{X}$-$M_\Delta$}

\citet{sta+al10} presented a computational study of the intrinsic covariance of cluster observables using the Millennium Gas Simulations. Two different physical treatments were proposed: shock heating driven by gravity only, or a second treatment with cooling and preheating. The predictions in \citet{sta+al10} on the scaling between mass and bolometric X-ray luminosity are strongly dependent on the adopted scheme. Acceptable values of the slope are in the range $1.1\ls \beta \ls 1.9$, consistent with our estimate of $\beta_{Y\textrm{-}Z}=1.8\pm0.3$. 

\citet{ett+al04b} noted a significant negative evolution in the $L_\mathrm{X}$-$M_\Delta$ relation ($\gamma_z \sim -0.94$) in a sample of local and high-redshift galaxy clusters extracted from a large cosmological hydrodynamical simulation with cooling and preheating, supporting the first evidence of a negative evolution (with respect to the self-similar model) observed in {\it Chandra} data of high-$z$ clusters in \citet{ett+al04a}. This negative evolution with $z$ is in agreement with our findings ($\gamma_z= -1.7\pm0.8$), even though the large statistical error make the estimate marginally compatible with self-similarity. We noted negative evolution in two different samples of X-ray clusters, i.e, the sample from \citet{mau+al12} and the MCXC, whose luminosities were largely based on independent data-sets and procedures.

A exhaustive study of the joint effect of feedback from supernovae (SNe) and active galactic nuclei (AGNs) on the evolution of the X-ray scaling laws presented in \citet{sho+al10} predicts an opposite behaviour of the evolution of the $L_\mathrm{X}$-$M_\Delta$ relation. They found that the energy output from SNe and AGNs as implemented through semi-analytic models of galaxy formation causes a positive evolution. On the other hand, simulations based on a pre-heating model where an entropy floor of 200~keV cm$^2$ is introduced at $z=4$ confirmed the presence of a negative evolution \citep{sho+al10}.

Positive evolution (i.e., higher luminosities at higher redshift, for a fixed mass, than the self-similar prediction) was also found in \citet{pik+al14} in a series of radiative hydrodynamical models, which suggested that radiative cooling is the main driver for departures from self-similarity. 

The pre-heating scenario seems to be preferred from our results.

\subsubsection{$Y_\mathrm{SZ}$-$M_\Delta$}

The consensus from numerical simulations is that the $Y_\mathrm{SZ}$-$M_\Delta$ is approximately self-similar in mass ($1.6\ls \beta \ls 1.8$) and it is characterised by a small intrinsic scatter \citep{sta+al10,kay+al12,bat+al12}. \citet{sta+al10} found that the evolution with redshift might be negative ($\gamma_z \sim -0.34$) in presence of cooling and preheating. 

In agreement with simulations, we did not detect any departure from the self-similar scaling, with $\beta_{Y\textrm{-}Z}=1.50\pm0.21$. The uncertainty in the observed time-evolution is too large ($\gamma_z= -0.8 \pm0.8$) to infer any statistically significant deviation from self-similarity. In fact, the evolution with mass is fully consistent with the findings of \citetalias{ser+al14_comalit_II}, where we found $\beta_{Y\textrm{-}Z}=1.37\pm0.15$ assuming a self-similar redshift evolution.

\subsubsection{$\sigma_\mathrm{v}$-$M_\Delta$}

Numerical simulations confirmed that the $\sigma_\mathrm{v}$-$M_\Delta$ relation is consistent with the self-similar scaling with mass \citep{evr+al08,mun+al13,sar+al13}. Some differences may arise from the galaxy population used to estimate the velocity dispersion and from the impact of selection using galaxy colour, projected separation from the cluster centre, galaxy luminosity, and spectroscopic redshift \citep{sar+al13}. Whereas dark-matter particles in simulations trace a relation that is fully consistent with the theoretical expectations, sub-haloes and galaxies trace slightly steeper relations with $\beta$ just above 1/3, and with slightly larger values of the normalisation \citep{mun+al13}. This is due to dynamical processes, namely dynamical friction and tidal disruption, which act on substructures and galaxies, but not on dark matter particles. The relevance of these effects depends on the halo mass and the effectiveness of baryon cooling, and may create a non-trivial dependence of the scaling relation on the tracer, the halo mass, and its redshift \citep{mun+al13}. A better statistical accuracy than that achieved with our results is needed to detect such effects.

\citet{sar+al13} noted a substantial agreement between the time evolution of the $\sigma_\mathrm{v}$-$M_\Delta$ relation and the expected self-similar evolution, $\gamma_z\sim 0$, which we confirm here within the statistical uncertainties.

The main sources of bias and scatter in velocity dispersion at fixed mass are the halo triaxiality, sampling noise, the presence of multiple kinematic populations within the cluster, and the effect of interlopers \citep{sar+al13}. \citet{sar+al13} found  $\sigma_{\log(\sigma_\mathrm{v}/M_\Delta)} \sim 0.05$ locally, and that the intrinsic scatter increases with redshift, with velocity dispersions that are $\sim$25 per cent less accurate for estimating single cluster masses at $z=1$ than at low redshift. \citet{sta+al10} found a smaller scatter of $\sigma_{\log(\sigma_\mathrm{v}/M_\Delta)} \sim 0.02$ for dark matter particles. 

Our findings, i.e., $\sigma_{\log(\sigma_\mathrm{v}/M_\Delta)} \sim 0.06\pm0.02$, slightly exceed the theoretical predictions. 
The accuracy of our results is not good enough to appreciate any evolution of scatter with redshift.

\subsection{Previous estimates}

We now discuss our findings in relation to previous results. We limited the comparison mainly to scaling relations which employed direct mass measurements based on weak lensing or the assumption of hydrostatic equilibrium, whereas we mostly discarded works based on mass estimates based on external calibration. Previous analyses of the $\sigma_\mathrm{v}$-$M_\Delta$ were mostly based on mass measurements assuming the dynamical equilibrium and the virial theorem \citep{gi+me01,rin+al13}. This mass measurement is direct too, but it is strongly correlated with the velocity dispersion, differently from the weak lensing masses we considered here.

We did not consider previous analyses of the conditional scaling relations in which the mass worked as the response variable. These relations are not easily inverted and cannot be compared straight on with our results.

\subsubsection{$L_\mathrm{X}$-$M_\Delta$}

Observed slopes of the $L_\mathrm{X}$-$M_\Delta$ relation from previous studies may be steeper than the self-similar prediction. Estimated values of $\beta$ ranges from 1.3 to 1.9 \citep{vik+al09,pra+al09,arn+al10,rei+al11,ett13}. Source of disagreement may be various. The slope and normalisation of the relation depend on the energy band and method used for the flux extraction \citep{ett15}. 

The intrinsic scatter of the $L_\mathrm{X}$-$M_\Delta$ relation is $\sim$40 per cent \citep{gio+al13}. It is the largest among the various X-ray scaling relations. We confirmed the large scatter in the $L_\mathrm{X}$-$M_\Delta$ relation. The X-ray luminosity is heavily affected by non-gravitational processes, the presence of cool-cores, and the overall dynamical state of the halo \citep{gio+al13}. Most of the scatter derives from the inner regions where cooling and merging effects are most pronounced. Our estimate of the scatter based on the the soft band luminosities provided by the MCXC is fully consistent with most of the previous results ($\sim40$ per cent), whereas the estimate based on the core-excised bolometric luminosities from \citet{mau+al12} is significantly smaller ($\sim20$ per cent). This suggests that a careful choice of the energy band and of the methods for the flux measurement might significantly reduce the intrinsic scatter of the $L_\mathrm{X}$-$M_\Delta$ relation.

A negative time-evolution of the relation was firstly noticed in \citet{ett+al04a} and later confirmed by \citet{rei+al11}, which found $\gamma_z=-1.3\pm0.2$ applying a tentative selection-bias correction. Our results confirm these previous findings.

From a multivariate analysis aimed to study X-ray luminosity, temperature, and gas mass fraction in a sample of clusters with well measured WL masses in the context of cosmological parameter determinations with cluster abundances, \citet{man+al14} estimated a slope of $1.71\pm0.17$ and an intrinsic scatter of $42\pm5$ per cent for the $L_\mathrm{X_{soft}}$-$M_\Delta$ relation assuming self-similar time evolution. Apart from the assumed redshift dependence, these results are directly comparable to ours, since \citet{man+al14} considered the scatter of the WL mass and the effects of the selection function. The estimated slope is in very good agreement with our result $\beta_{Y|Z}=1.60\pm0.27$. In principle, slopes obtained from a multivariate analysis may differ from the results of a single $O$-$M_\Delta$ relation if the considered observables are strongly correlated \citep{ett13}. However, this is not the case of the X-ray properties considered in \citet{man+al14}. 

\citet{roz+al14c} developed a self-consistent method to derive scaling relations satisfying optical data from SDSS, X-ray data from ROSAT and {\it Chandra}, and SZ data from {\it Planck}. Assuming a self-similar time-evolution, they derived a slope of $\beta_{Y|X}=1.55\pm 0.09$ for the $L_\mathrm{X_{soft}}$-$M_{500}$ relation with a scatter of $\sim39\pm3$ per cent. Slope and scatter from \citet{roz+al14c} are consistent with our results for the $L_\mathrm{X_{soft}}$-$M_\Delta$ relation based on the MCXC (see Sec.~\ref{sec_resu}) even though the analyses presents some major differences. In fact, \citet{roz+al14c} used stacked data rather than measurements from single clusters and they assumed a self-similar time evolution.

\subsubsection{$Y_\mathrm{SZ}$-$M_\Delta$}

Observed slopes of the scaling relation between mass and SZ flux are discordant to some degree \citepalias{ser+al14_comalit_II}. The {\it Planck} team determined the $Y_\mathrm{SZ}$-$M_{500}$ relation relying on masses estimates based on  the $Y_\mathrm{X} $ proxy and assuming a self-similar time-evolution \citep{planck_2013_XX}. Through the BCES-orthogonal regression, they found $\beta_{Y\textrm{-}X}=1.79\pm0.06$  and an intrinsic orthogonal scatter $\sim$ 15-20 per cent \citep{planck_2013_XX}. A previous calibration  based on 19 weak lensing clusters mainly from the LoCuSS sample \citep[Local Cluster Substructure Survey,][]{oka+al10} gave $\beta_{Y\textrm{-}X} = 1.7 \pm 0.4$ \citep{planck_int_III}. However, weak lensing masses of the LoCuSS clusters are biased low due to contamination effects and systematics in shape measurements \citep{oka+al13}. The underestimate might be mass dependent and affect the estimated slope. With a self-consistent method, \citet{roz+al14c} found a slope of $\beta_{Y|X}=1.71\pm 0.08$ with a scatter of $\sim15\pm2$ per cent. These estimated slopes are steeper but consistent within the statistical uncertainty with our result of $\beta_{Y\textrm{-}X}=1.5\pm0.2$.

\citet{and14} claimed that the evolution of the $Y_{SZ}$-$M_{500}$ relation is significantly inconsistent with the self-similar evolution. He found $\gamma_z=1.8\pm0.4$. The disagreement with our result, which is fully consistent with the self-similar prediction, might be due to the arbitrary choice in \citet{and14} to use a pre-determined completeness function. An inappropriate modelling can bias the estimates of the scaling relations as well as neglecting time-evolution at all.

Furthermore, \citet{and14} neglected the intrinsic scatter in the mass estimate, and, following \citet{planck_2013_XX}, he employed a mass proxy based on $Y_\mathrm{X} $, which is strongly correlated to the SZ signal. These choices likely explain the disagreement with our result.

\subsubsection{$\lambda$-$M_\Delta$}

Most of the analyses correlating optical richness to mass were based on stacking techniques \citep[ and references therein]{roz+al14c,cov+al14}. Here, we focus on not binned data. \citet{wen+al09} considered a compilation of clusters whose masses had been estimated by X-ray or weak-lensing methods to infer a slope of $\beta_{Y\textrm{-}Z}=1.17\pm0.03$.

\citet{an+co14} studied the mass--richness scaling for a sub-sample of the CCCP clusters \citep[The Canadian Cluster Comparison Project,][]{hoe+al12}. They found that richness scales almost linearly with the projected weak-lensing mass ($\beta_{Y|X}=1.3\pm0.3$), with a statistically insignificant evolution ($\gamma_z= -0.7\pm0.7$). The evolution with mass measured in \citet{an+co14} is steeper than a previous result by \citet{an+be12}, which used spherical masses and projected richnesses to find $\beta_{Y|X}=0.46\pm0.12$. 

The slower slope found in \citet{an+be12} might be due to their assumption that the masses, which were measured with the caustic method, were unscattered measurements of the true masses. However, masses based on this method are highly scattered \citepalias{ser+al14_comalit_II}. Neglecting this effect makes relations flatter \citepalias{se+et14_comalit_I,ser+al14_comalit_II}. The very large scatter of $\sim58\pm7$ per cent measured in \citet{an+be12} is biased high for the same reason.

\section{Conclusions}
\label{sec_conc}

To assess the role of the evolution with time of the scaling relations between mass and observed properties it is crucial to avoid biases in the calibration. The evolution of cluster scaling relations is still debated. Small samples or biases due to the (unknown) selection function are two of the main problems. We developed a regression methodology that at the same time constrains the evolution and calibrates the completeness function of the studied sample. 

The method is general and lets the data determine the time-dependent scatter of the mass proxy or the intrinsic scatter between the observable and the mass. Selection functions and their time-dependence are often not known a priori. In our approach, the selection function can be determined in the context of the regression procedure. The approach we implemented is Bayesian and can easily include any additional or a priori information on the completeness of the sample. This method is functional in the context of large photometric surveys such as Euclid \citep{eucl_lau_11}, where self-calibration of scaling relations is crucial to unbiased estimates of dark energy with study of cluster abundances \citep{ma+mo04}

We tested the method with large heterogeneous samples to calibrate either optical properties, such as richness and velocity dispersion, or observables connected to the intra-cluster medium, such as X-ray luminosity and SZ flux. Masses were estimated with weak lensing analyses and intrinsic scatter in the mass estimate was considered. Weak lensing masses are reliable mass measurements up to high redshifts. 

To our knowledge and not considering mass estimates based on the viral theorem, this is one of the first studies to compare galaxy velocity dispersions to direct estimates of masses of clusters (weak lensing masses in our case). This is the approach usually followed in numerical simulations \citep{evr+al08,mun+al13,sar+al13} to built mass proxies based on the velocity dispersion without assuming dynamical equilibrium and without exploiting the properties of the infall patterns.

We found that observables scale self-similarly with respect to the mass and that they evolve self-similarly with cosmic time. The only exception is the $L_\mathrm{X}$-$M_\Delta$ relation, which seems to show a negative evolution. A similar level of evolution can be obtained in hydrodynamical simulations of the intracluster medium by including additional radiative cooling and uniform preheating at high redshift as a simple model of non-gravitational heating from astrophysical sources \citep{ett+al04b,sho+al10,rei+al11}. 

The intrinsic scatter in the mass--velocity dispersion relation is notably small, which encourages the use of velocity dispersions as mass proxies. Our procedure accounts for time evolution of the intrinsic scatter. However, the large statistical uncertainties made the parameters characterising this dependence as de facto noise parameters which had to be marginalised over to avoid to bias low the uncertainties on the scaling relation. The determination of this evolution is out of reach in present samples of nearly one hundred of clusters  \citep{man+al10b}, but is should be feasible in future large surveys that will be able to collect homogeneous multi-band information of an unprecedented number of clusters.

\section*{Acknowledgements}
M.S. thanks Adam Mantz, Kenneth Rines, and Alex Saro for clarifying some aspects of their works. M.S. acknowledges financial contributions from contracts ASI/INAF n.I/023/12/0 `Attivit\`a relative alla fase B2/C per la missione Euclid', PRIN MIUR 2010-2011 `The dark Universe and the cosmic evolution of baryons: from current surveys to Euclid', and PRIN INAF 2012 `The Universe in the box: multiscale simulations of cosmic structure'. S.E. acknowledges the financial contribution from contracts ASI-INAF I/009/10/0 and PRIN-INAF 2012 `A unique dataset to address the most compelling open questions about X-Ray Galaxy Clusters'. This research has made use of NASA's Astrophysics Data System (ADS) and of the NASA/IPAC Extragalactic Database (NED), which is operated by the Jet Propulsion Laboratory, California Institute of Technology, under contract with the National Aeronautics and Space Administration.


\setlength{\bibhang}{2.0em}

\appendix

\section{The mass function}
\label{app_mass}

\begin{figure}
       \resizebox{\hsize}{!}{\includegraphics{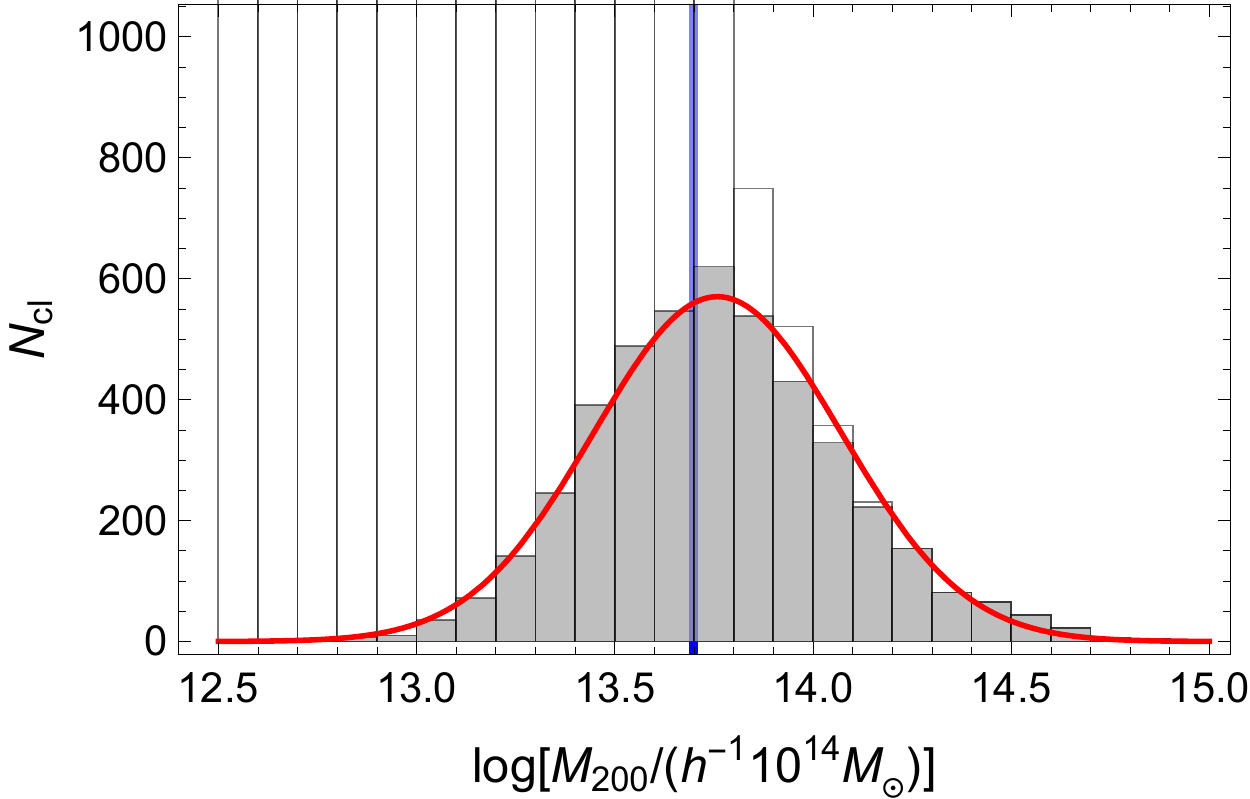}}
       \caption{Distribution of true masses $M_{200}$ of a sample of clusters at $z=0.3$ (grey histogram) whose observable proxy mass was larger than $M_{\mathrm{th},200}=0.5 \times 10^{14} M_\odot/h$ (vertical blue line). The assumed log-normal scatter is $\sigma=0.25$. The empty histogram represents the full distribution of the true masses of all clusters before the selection; the grey histogram represent the distribution of selected clusters, i.e, clusters whose observed proxy mass is above the threshold (vertical blue line); the red line is the Gaussian approximation to the mass function.. The true masses were extracted from a cosmological halo mass function following \citet{tin+al08}.}
	\label{fig_pM200_Tinker}
\end{figure}

\begin{figure}
       \resizebox{\hsize}{!}{\includegraphics{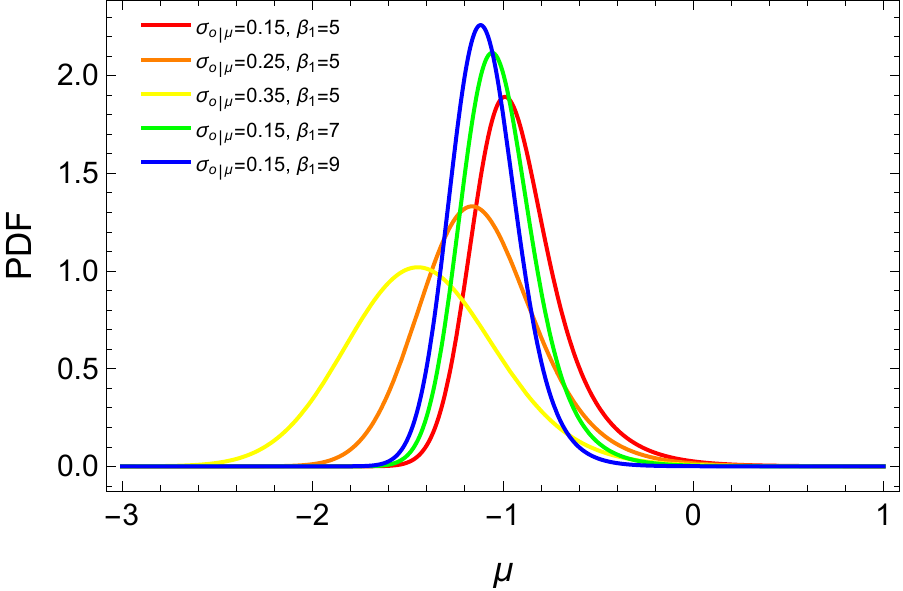}}
       \caption{Probability function of the log-mass $\mu$, i.e., the logarithm of the over-density mass $M_\Delta$. We assume that $\mu$ is intrinsically distributed as a local power-law and that clusters are selected if they have a proxy mass $o>-1$. The value of the slope of the halo function $\beta_1$, and of the intrinsic scatter for the different cases are reported in the legend.}
	\label{fig_pM_Gaussian}
\end{figure}

The mass distribution of an observationally selected sample of clusters is usually limited at large masses by the steepness of the mass function, and, at smaller masses, it is limited by some observational thresholds they have to surpass to be detected/selected. This two factors conjure to make the distribution of the masses approximately log-normal \citep{li+hu05}. 

As an example, we extracted a number of clusters from the cosmological halo mass function \citep{tin+al08}. The clusters were then selected if their observable proxy mass was in excess of a given threshold value. We assumed that the proxy masses were unbiased, i.e., the expected value of the proxy for a given mass is exactly the mass, but (log-normally) scattered with respect to the true masses. The final distribution, which is plotted in Fig.~\ref{fig_pM200_Tinker}, resembles a normal function.

The mass function of an observed sample can be expressed as
\beq
\label{eq_pmu_1_0}
\frac{dn_\mathrm{sample}}{d \mu}= \chi (\mu) \frac{dn}{d \mu},
\eeq
where $\mu \equiv \log M_\Delta$ and $dn/d\mu$ is the number density per logarithmic interval. The completeness of the observed sample $\chi$ can be usually approximated as
\beq
\label{eq_pmu_1_1}
\chi (\mu) \simeq \frac{1}{2}  \mathrm{erfc} \left( \frac{\mu_\mathrm{\chi}-\mu}{\sqrt{2}\sigma_\mathrm{\chi}}\right),
\eeq
where $\mathrm{erfc}$ is the complementary error function.

Some considerations based on a toy-model can give us a better insight. The mass function in small mass and redshift ranges can be approximated using a first-order Taylor expansion as \citep{roz+al14b,evr+al14},
\beq
\label{eq_pmu_1_2}
\frac{dn}{d \mu} \propto \exp \left( -\beta_1 \mu \right).
\eeq
At large masses, the mass function is steep and the function in Eq.~(\ref{eq_pmu_1_2}) provides a good approximation. Fitting the approximation to the mass function proposed by \citet{tin+al08}, we found that at $z\simeq 0.3$ (1.0), $\beta_1 \simeq 6.8$ (11.3) around a pivot mass of $M_{200}=5 \times10^{14}M_\odot/h$ for the standard $\Lambda$CDM cosmology. 

Previous treatments have approximated the mass function of the sample with the cosmological halo distribution in Eq.~(\ref{eq_pmu_1_2}) \citep[ and references therein]{ma+se07,all+al11,roz+al14b,evr+al14}. Here we want to consider the additional effect of the selection function, which severely limits the number of low mass halos. Let us assume that the observable $O$ is log-normally scattered around the mass. The conditional probability for the log-variable $o \equiv \log O$ is
\beq
\label{eq_pmu_2}
p (o | \mu) = \frac{1}{\sqrt{2 \pi} \sigma_{o|\mu}} \exp \left[ -\frac{1}{2}\left( \frac{o-\mu}{\sigma_{o|\mu}} \right)^2\right].
\eeq
To simplify the notation, we assumed that $O$ scales linearly with the mass and that is was normalised so that the expected value of $o$ for a given mass is exactly $\mu$.

If we select a cluster sample imposing a hard cut in the observable, $o > o_\mathrm{th}$, the resulting mass function of the selected clusters is
\beq
\label{eq_pmu_3}
p(\mu)= \frac{\beta_1}{2}  \mathrm{erfc} \left( \frac{o_\mathrm{th}-\mu}{\sqrt{2}\sigma_{o|\mu}}\right)  \exp \left[ -\beta_1 \left( \mu - o_\mathrm{th}+ \frac{\beta_1}{2}\sigma_{o|\mu}^2 \right) \right].
\eeq
By comparison with Eq.~(\ref{eq_pmu_1_0}), we see that in this case the completeness function is exactly given by the complementary error function with $\mu_\chi=o_\mathrm{th}$ and $\sigma_\chi=\sigma_{o|\mu}$.

The probability distribution of the observable is
\beq
\label{eq_pmu_4}
p(o) = \beta_1 \exp \left[ -\beta_1 \left(o - o_\mathrm{th} \right) \right],
\eeq
for $o >o_\mathrm{th}$, and $p(o)=0$ below the threshold. The two-dimensional distribution is
\begin{multline}
\label{eq_pmu_5}
p(\mu,o)=  \beta_1 \frac{1}{\sqrt{2 \pi} \sigma_{o|\mu}} \exp \left[ -\frac{1}{2}\left( \frac{o-\mu}{\sigma_{o|\mu}} \right)^2\right] \\
\times \exp \left[ -\beta_1 \left( \mu - o_\mathrm{th}+ \frac{\beta_1}{2}\sigma_{o|\mu}^2 \right) \right],
\end{multline}
for $o > o_\mathrm{th}$, and it is null otherwise.

The mass function of the selected clusters in Eq.~(\ref{eq_pmu_1_0}), and its approximation in Eq.~(\ref{eq_pmu_3}) can be adequately described by a Gaussian distribution. The effectiveness of the Gaussian structural model for estimating the regression parameters was illustrated by \citet{kel07}, who showed that as far as the distribution of the covariate variable, i.e., the mass function in our case, is fairly unimodal a simple Gaussian can perform competitively with a mixture. For a large range of scatters, $0.05\ls \sigma_{\chi} \ls 0.25$, and slopes, $5 \ls \beta_1 \ls 20$, we found the following analytical approximation:
\beq
\label{eq_pmu_6}
p(\mu) \simeq \frac{1}{\sqrt{2 \pi} \sigma_\mu} \exp \left[ -\frac{1}{2}\left( \frac{\mu -\bar{\mu}}{\sigma_\mu} \right)^2\right].
\eeq
with
\beq
\label{eq_pmu_7}
\bar{\mu} \simeq \mu_\mathrm{\chi} +\sigma_{\chi} -1.21 \beta_1 \sigma_{\chi}^2,
\eeq
and
\beq
\label{eq_pmu_8}
\sigma_\mu \simeq 0.062 - 0.0023 \beta_1 + (1-0.0052\beta_1)\sigma_{\chi} ,
\eeq
Equations~(\ref{eq_pmu_7}) and~(\ref{eq_pmu_8}) can be used to approximately estimate the completeness function, Eq.~(\ref{eq_pmu_1_1}), if we know the mass distribution of the clusters, Eq.~(\ref{eq_pmu_6}). The larger the intrinsic scatter $\sigma_{\chi}$ and the steeper the mass function, the better the Gaussian approximation, see Fig.~\ref{fig_pM_Gaussian}. Even for small scatters ($\sigma_{o|\mu} \sim 0.15$) and shallow mass functions ($\beta_1 \sim 5$), the normal approximation is reliable.

\section{Catalogs of velocity dispersion}
\label{app_SC}

We compiled two catalogs of galaxy clusters with measured velocity dispersion, the Sigma Catalogs (SCs). SC-{\it all} comprises the full body of information. Multiple entries are present. The SC-{\it single} is a subsample with unique entries. When a cluster had multiple analyses available in literature, we picked for the SC-{\it single} the results based on the larger number of identified member galaxies with confirmed spectroscopic redshift, $N_\mathrm{members}$.

In each catalog, objects are ordered by right ascension. The format of the catalogs is as follows.
\begin{description}
\item Cols. 1-2: name of cluster as designated in the original paper.
\item Cols. 3-4: right ascension RA (J2000) and declination DEC (J2000), as quoted in the original paper. If coordinates are not quoted in the source paper or in a companion one, I reported the coordinates of the NED's association.
\item Col. 5: redshift $z$, as reported in the original paper.
\item Col. 6: external validation through NED. `N': the NED's object was associated by name; `P': the NED's object was associated by positional matching; `NA': no found association.
\item Cols. 7-11: as in cols. 1-5, but for the NED's association.
\item Col. 12: author code.
\item Col. 13: bibliographic code from NASA's Astrophysics Data System (ADS).
\item Col. 14: $N_\mathrm{members}$, number of confirmed member galaxies used to measure the velocity dispersion. If the number was not available, we put the entry to $-99$.
\item Col. 15: aperture radius within which member galaxies were looked for, in units of $\mathrm{Mpc}/h$. The radius is measured in a flat $\Lambda$CDM model with $\Omega_\mathrm{M}=0.3$. If the information is not available, we put the entry to $-99$.
\item Col. 16: line-of-sight velocity dispersion $\sigma_\mathrm{v}$, in units of km/s.
\item Col. 17: uncertainty on the line-of-sight velocity dispersion, in units of km/s, as quoted in the reference paper, $\delta_{\sigma_\mathrm{v,ref}}$.
\item Col. 18: Standardised uncertainty in the velocity dispersion $\sigma_\mathrm{v}$, in units of km/s, measured as
\beq
\delta_{\sigma_\mathrm{v,stand}}= \frac{0.92\sigma_\mathrm{v}}{\sqrt{N_\mathrm{members}-1}},
\eeq
or fixed to -99 if $N_\mathrm{members}$ is unknown.
\end{description}

The format of columns 1-13 follows the LC$^2$ \citepalias{ser14_comalit_III}. The catalogs are publicly available at \url{http://pico.bo.astro.it/\textasciitilde sereno/CoMaLit/sigma/} and they will be periodically updated.

\end{document}